\documentclass{vldb}
\usepackage{graphicx}
\usepackage{balance}  % for  \balance command ON LAST PAGE  (only there!)
\usepackage{hyperref}

\begin{document}

% ****************** TITLE ****************************************

\title{Targeting {\ttlit HIV-related} Medication Side Effects and Sentiment Using Twitter Data}

\author{Cosme Adrover, Ph.D.\titlenote{Corresponding author. For a detailed description of the analysis presented in this paper, including relevant code, please visit \href{http://www.cosme-adrover.com}{cosme-adrover.com}. Contact email: cosme.adrover.pacheco@gmail.com} \qquad Todd Bodnar \qquad Marcel Salath\'e, Ph.D. \vspace{4mm} \\ Center for Infectious Disease Dynamics\\Pennsylvania State University\\208 Mueller Lab\\University Park, PA 16802, U.S.A.}
% There's nothing stopping you putting the seventh, eighth, etc.
% author on the opening page (as the 'third row') but we ask,
% for aesthetic reasons that you place these 'additional authors'
% in the \additional authors block, viz.
%\additionalauthors{Additional authors: John Smith (The Th{\o}rv\"{a}ld Group, {\texttt{jsmith@affiliation.org}}), Julius P.~Kumquat
%(The \raggedright{Kumquat} Consortium, {\small \texttt{jpkumquat@consortium.net}}), and Ahmet Sacan (Drexel University, {\small \texttt{ahmetdevel@gmail.com}})}
\date{\today}
% Just remember to make sure that the TOTAL number of authors
% is the number that will appear on the first page PLUS the
% number that will appear in the \additionalauthors section.

\maketitle

\begin{abstract}
We present a descriptive analysis of Twitter data. Our study focuses on extracting the main side effects associated with HIV treatments. The crux of our work was the identification of personal tweets referring to HIV. We summarize our results in an infographic aimed at the general public. In addition, we present a measure of user sentiment based on hand-rated tweets. 
\end{abstract}

\section{Introduction}
Twitter is a popular micro-blogging platform where users share thoughts and emotions. Everyday, hundreds of millions of tweets are posted on Twitter. This offers a large potential source of information for translational medicine, like for instance, to predict regions with large likelihood disease outbreaks based on tweeted vaccination sentiments~\cite{marcel:2011}. Moreover, this large pool of available data can be used to estimate what the most common side effects of medication are, as quoted by online users. 

\vskip 2mm
The study that we present is based upon tweets filtered by specific keywords related to HIV and HIV treatments. Our goal was to determine if there are common side effects related to a particular HIV treatment, and to establish overall user sentiment.

\vskip 2mm
Through our study, we defined methods to target populations of interest. We used crowdsourcing Amazon Mechanical Turk to rate tweets to create training samples for our machine learning algorithms. On the analytical side, these algorithms were used to identify our targeted community: users with HIV whose tweets included references to treatments, symptomatic descriptions, opinions, feelings, etc.

\section{Datasets}
We used tweets collected between September 2010 and August 2013. These tweets were filtered by at least one of the following keywords contained within the tweet: 
Sustiva, Stocrin, Viread, FTC, Ziagen, 3TC, Epivir, Retrovir, Viramune, Edurant, Prezista, Reyataz, Norvir, Kaletra, Isentress, Tivicay, Atripla, Trizivir, Truvada, Combivir, Kivexa, Epzicom, Complera, Stribild, HIV treatment, HIV drug, anti-hiv, triple therapy hiv, anti hiv. The sample size is of 39,988,306 tweets.

\section{Processing of tweets}\label{processing}
For each collected tweet, we created a list of all the distinct tokens contained in the tweet. If at least one keyword matched at least one item in the list, we kept that tweet. Otherwise, we discarded the tweet. This step reduced our data sample to about 1.8 million tweets, mainly due to the presence of compound words such as \emph{giftcard}, triggered by the keyword FTC, that represented a large source of noise. Moreover, a subsample of tweets containing the keyword FTC presented a large bias towards information related to the \emph{Federal Trade Commission}. We decided to discard this subsample for this study. After this further reduction of the data, the sample to analyze contained 316,081 tweets.

\vskip 2mm
Based on random trials of human-rated tweets, we discarded tweets containing the following words: \emph{bit.ly, t.co, million, http, free, buy, news, de, e, za, que, en, lek, la, obat, da, majka, molim, hitno, mil, africa}; also the tweets starting with \emph{HIV}, tweets containing \emph{3TC} and that were posted by users identified as non-English speakers on Twitter. We estimate that this removal lead to a maximum loss of about 300 tweets from our targeted community.

\vskip 2mm
Following the aforementioned processing of tweets, the sample was drastically reduced to 37,337 tweets, about 0.1$\%$ of the original sample size. In the next section, we describe how we identified tweets posted by our community of interest.

\section{Identification of targeted community}
\begin{figure}
\centering
\includegraphics[height=2in]{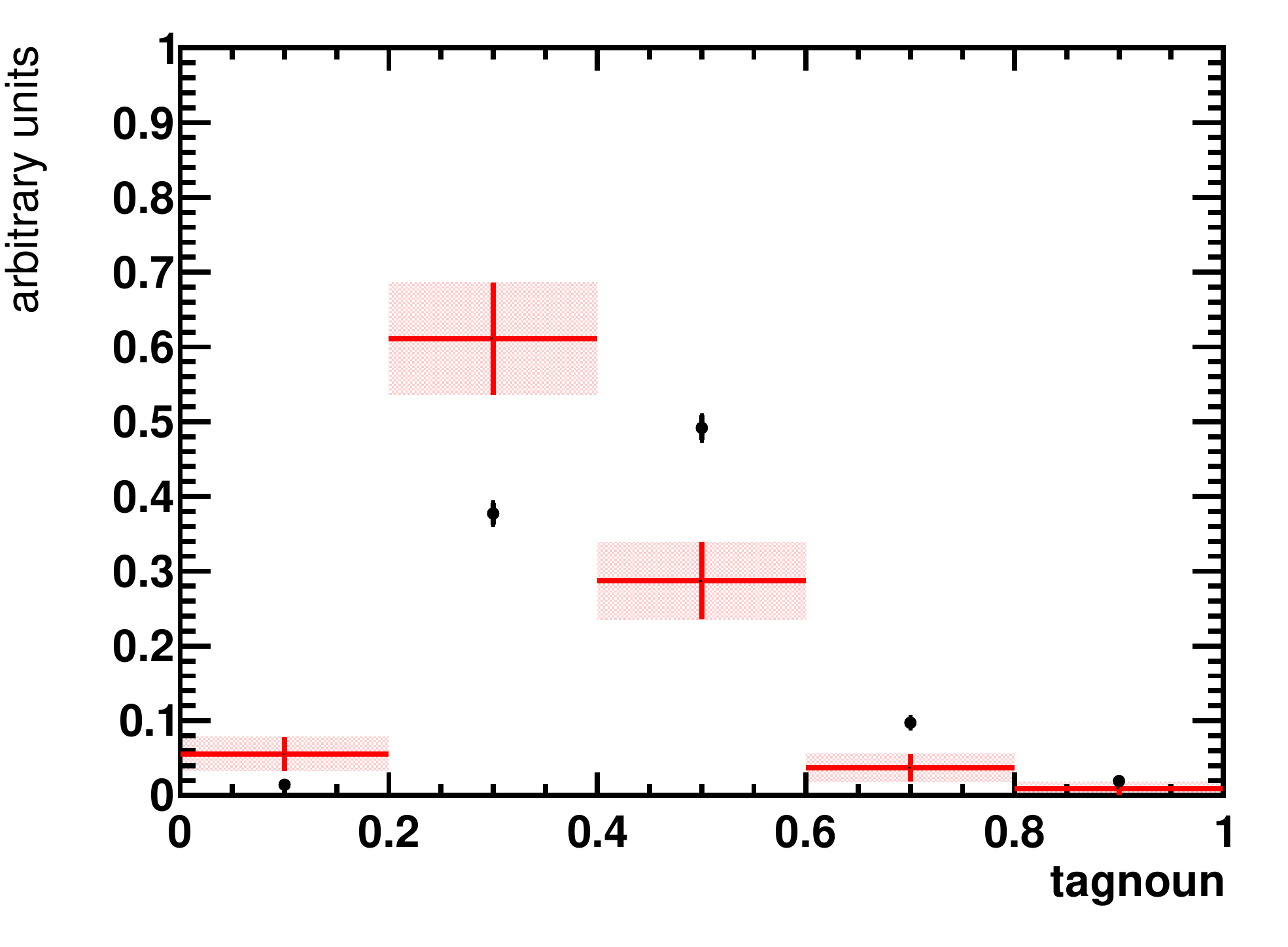}\\
\includegraphics[height=2in]{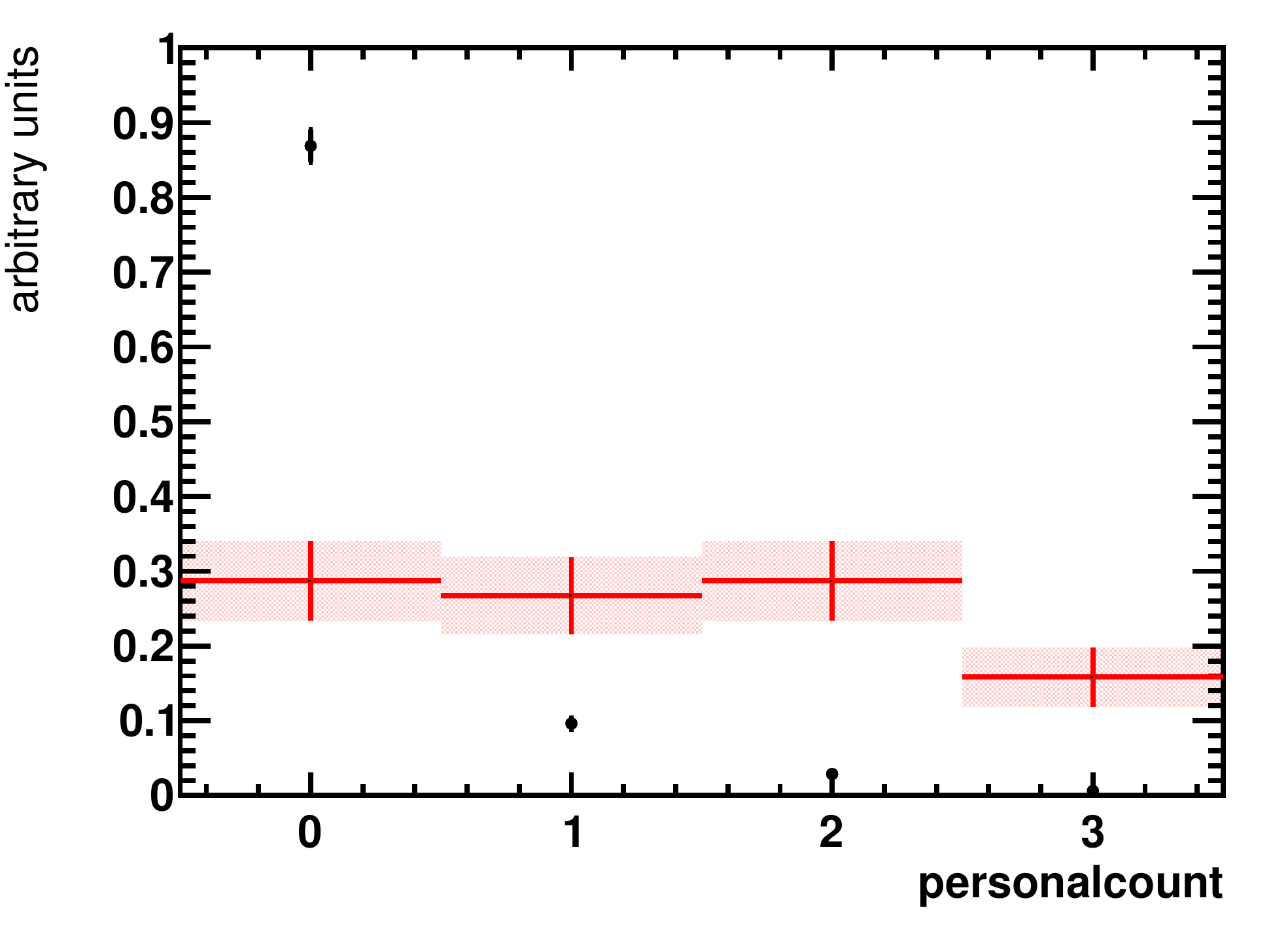}\\
\caption{Distributions of tagnoun (top) and personalcount (bottom) for noise (black dots) and targeted community (red). The statistical uncertainties are represented on the vertical axis. The horizontal red lines are bins and not uncertainties.}
\label{variables}
\end{figure}

Our goal was to be able to identify the targeted community of users making personal posts about HIV from the remaining noise contained within our data. This noise mainly constitutes tweets characterized by an objective message, such as news regarding a particular HIV treatment. To remove the noise and attain our goal, we defined a set of features that aimed to transform the tweet into quantitative information. Moreover, these features were selected based on their separation power between objective information and subjective sentences charged with personal references. Tab.~1 summarizes the description of each feature used in our analysis. Fig.~\ref{variables} displays two of these features obtained using crowdsourcing rated data for both targeted community and noise. This figure shows how these features are indeed useful tools to distinguish targeted community from noise, our main goal in this section~\cite{featuremine}.

\begin{table*}\label{features}
\centering
\caption{List of features used in our analysis.}
\begin{tabular}{|l|l|} 
\hline
    Feature & Definition \\
\hline
\hline
personalcount & Number of first person pronouns in tweet. \\
is\_notenglish & Number of words contained in a list of most common words in tweets rated as not english. \\
\hline
\hline
tagnoun  & Number of nouns in tweet divided by the total number of words. \\
\hline
\hline
sis\_noise & Ratio of a measure of similarity between tweet and noise corpus by its uncertainty.\\
sis\_signal & Ratio of a measure of similarity between tweet and targeted community corpus by its uncertainty.\\
\hline
\hline
bigrams\_noise & Weighted number of bigrams in tweet contained in corpus of noise tweets.\\
\hline
\hline
is\_english & Ratio of number of words in English by number of words in foreign languages.\\
\hline
\hline 
common\_noise & Weighted number of words in tweet similar to common words in noise corpus.\\
common\_signal & Weighted number of words in tweet similar to common words in targeted community corpus.\\
\hline
\hline
wordscount & Number of words in tweet. \\
ncharacters & Number of characters in tweet.\\
\hline
\end{tabular}
\end{table*}

\vskip 2mm
At the same time, we aimed to remove non-English tweets. The method that we proposed applies the following requirements: \emph{is\_english}$\geq$1, \emph{ncharacters}$<$150, \emph{in\_notenglish}$<$14\\ and $\frac{\emph{wordscount}}{\emph{in\_notenglish}}>1$. We estimated that these requirements lead to a 6$\%$ loss of targeted tweets, while removing 20$\%$ of all remaining noise and 94$\%$ of non-English tweets.

\begin{figure}
\centering
\includegraphics[width=2.5in,height=2in]{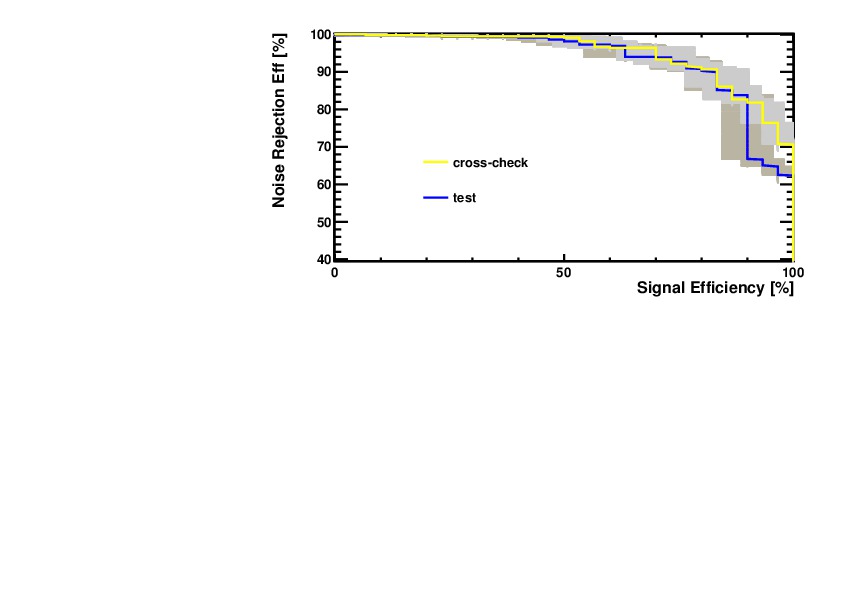}
\caption{Classifier efficiency on targeted tweets (Signal) versus noise rejection efficiency for two different testing samples. Gray bands represent statistical uncertainties.}
\label{comparison}
\end{figure}

\vskip 2mm
We utilized the TMVA library~\cite{tmva} to define our machine learning classifier. A Support Vector Machines classifier, trained using the variables personalcount, tagnoun, sis\_noise, sis\_signal, bigrams\_noise, is\_english, common\_noise,\\ common\_signal and ncharacters, allowed to reduce the noise by 80$\%$ while keeping 90$\%$ of the targeted tweets (Fig.~\ref{comparison}). We refer to this point on the ROC curve as it represents our working region. This region allows us to keep a large fraction of interesting tweets, while removing the necessary noise at the level to permit crowdsourcing rating that is within our budget. 

\vskip 2mm
We were confident with the performances of our classifier as the results obtained using a different testing (cross-check) sample agree within the statistical uncertainties (Fig.~\ref{comparison}). Tab.~2 summarizes the data sample sizes used for training, testing and validating our classifier. We tested other algorithms such Boosted Decision Trees and Artificial Neural Networks, but Support Vector Machines outperforms the former two in our region of interest. 

\begin{table}\label{yields}
\centering
\caption{Number of tweets used to train, test and validate our classifier, for both noise and targeted HIV-related tweets.}
\begin{tabular}{|l|l|l|l|} 

\hline
 Data Type    & Training & Testing & Validation \\
\hline
\hline
Noise &  603 & 603 & 603 \\
\hline
Targeted & 49 & 30 & 30 \\
\hline
\end{tabular}
\end{table}

\vskip 2mm
After using the trained classifier to reduce our sample of 37,337 tweets we obtained 5,543. These tweets were then sent for crowdsource rating, leading to 1,642 tweets from our targeted community, posted by 518 unique users. Taking into account analysis requirements, this last figure of tweets approximately represents 75$\%$ of the starting targeted community. 

\vskip 2mm
For a more a detailed version of this section, as well as relevant code, more information is available in the Appendix and at \href{http://www.cosme-adrover.com}{cosme-adrover.com}.

\section{Analysis of targeted community tweets}

\begin{figure}
\centering
\includegraphics[width=3.5in,height=2.5in]{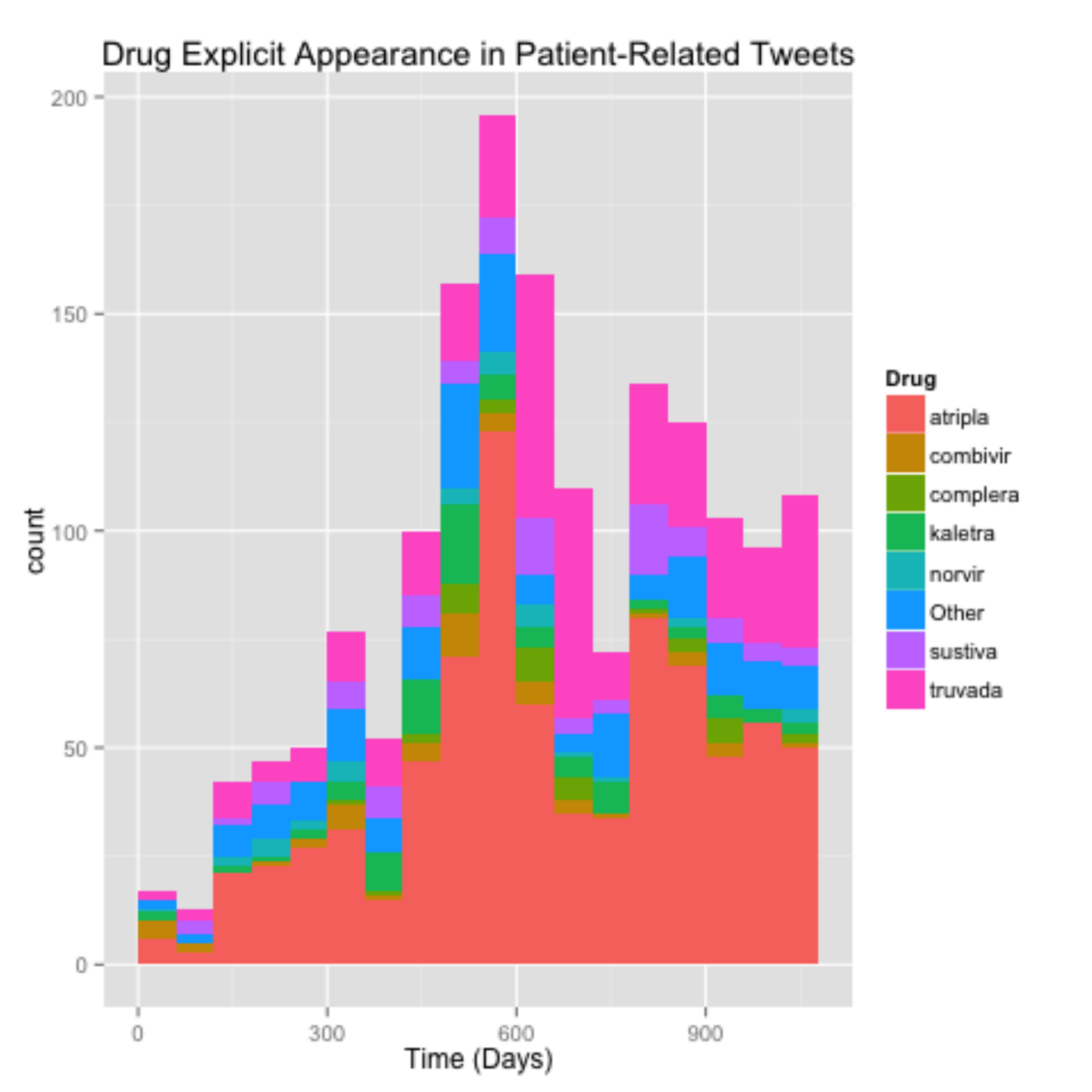}
\caption{Drug mentions from 09/09/2010 until 08/28/2013. Each bin spans a total of 60 days.}
\label{drugstime1}
\end{figure}

In the previous section we described how we identified users that tweet about their daily lives in the context of HIV. We present in Fig.~\ref{drugstime1} a visualization of HIV drugs mentioned by users from 09/09/2010 through 08/28/2013. In this plot, made using ggplot~\cite{ggplot}, we depict the seven most mentioned drugs separately, and group the rest under the label '\emph{Other}'. Fig.~\ref{drugstime1} presents a peak during the first semester of 2012. The total number of tweets is shared among all drugs equally in the first two bins. This trend disappears after the third bin, where atripla receives more explicit mentions. The only period where Atripla$\copyright$ is not ranked first in terms of mentions corresponds to the time spanning from May through August 2012, when Truvada$\copyright$ gained in popularity.

\begin{figure}
\centering
\includegraphics[width=3.5in,height=2.5in]{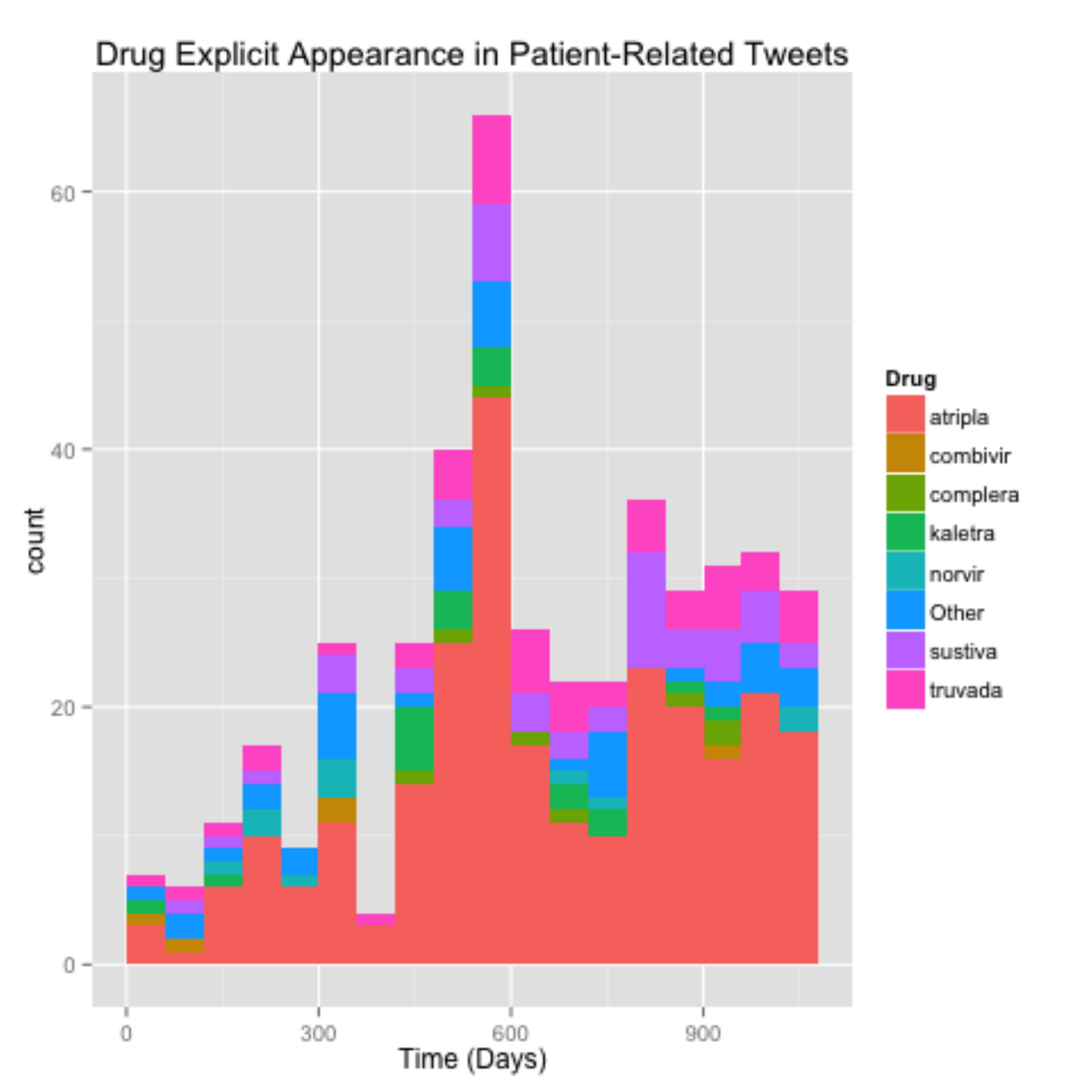}
\caption{Drug mentions from 09/09/2010 until 08/28/2013 in tweets that are not re-tweets and have mentions to side effects. Each bin spans a total of 60 days.}
\label{drugstimenortsides}
\end{figure}

\vskip 2mm
In Fig.~\ref{drugstimenortsides} we study drug appearance over time for tweets that are not re-tweets and that contain references to side effects. These side effects were hand-rated. Fig.~\ref{drugstimenortsides} presents a clear peak during March and April 2012. Moreover, we see that the mentions to Truvada$\copyright$, used as part of a strategy to reduce the risk of HIV infection, are highly reduced when compared to Fig.~\ref{drugstime1} Spring-Summer 2012 period. 

\begin{figure}
\centering
\includegraphics[width=3.5in,height=2.5in]{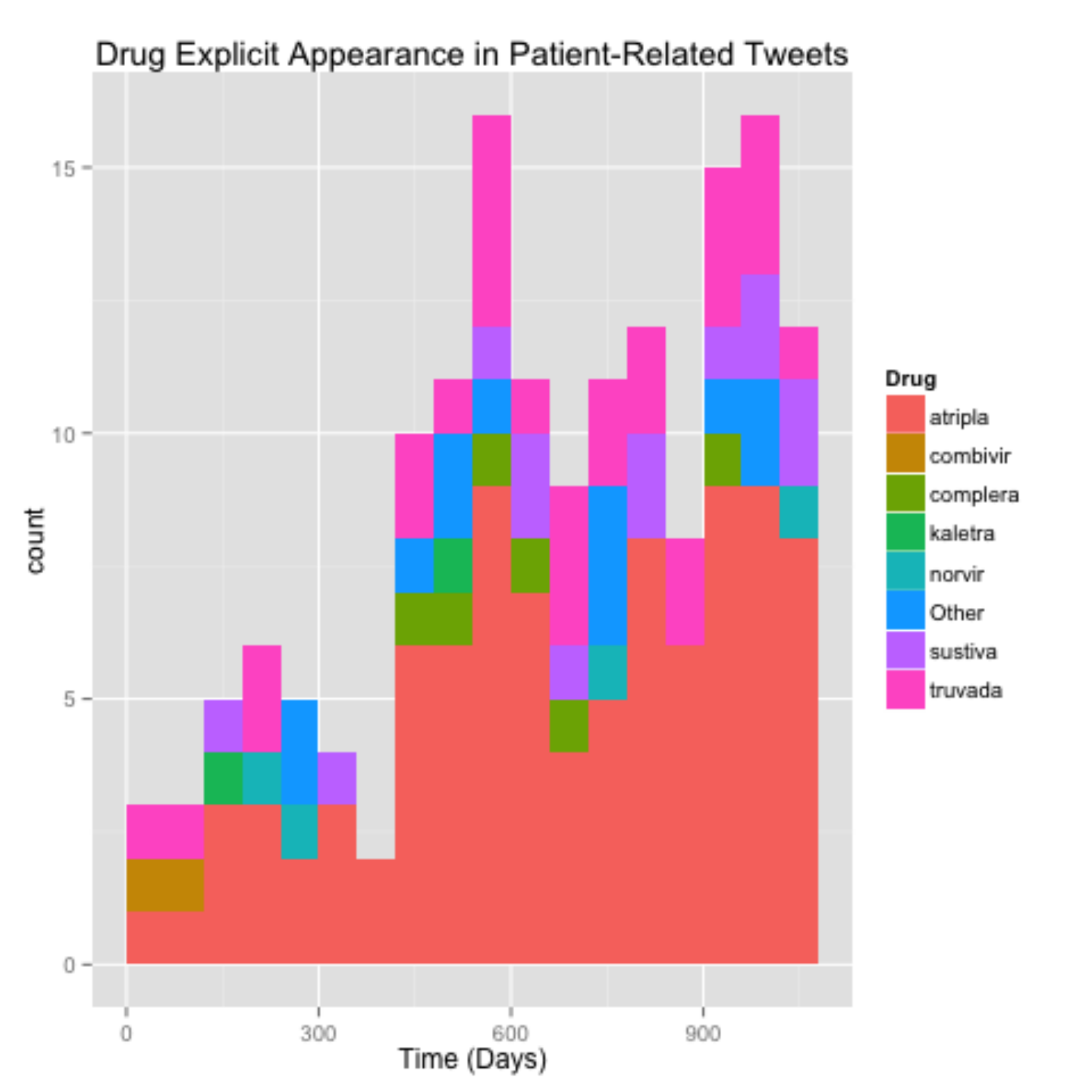}
\caption{Drug mentions from 09/09/2010 until 08/28/2013. Each bin spans a total of 60 days. Retweets are excluded, only tweets with mentions to side effects and from unique users over time are considered.}
\label{drugstimenortsidesunique}
\end{figure}

\vskip 2mm
We investigate possible reasons for the peaks in Figs.~\ref{drugstime1},~\ref{drugstimenortsides} and conclude that one possible explanation relies on the fact that many new users shared their side effects for the first time during this period (Fig.~\ref{drugstimenortsidesunique}).

\subsection{Study of Side Effects}
\begin{figure*}
\centering
\includegraphics[width=9in,angle=90.1]{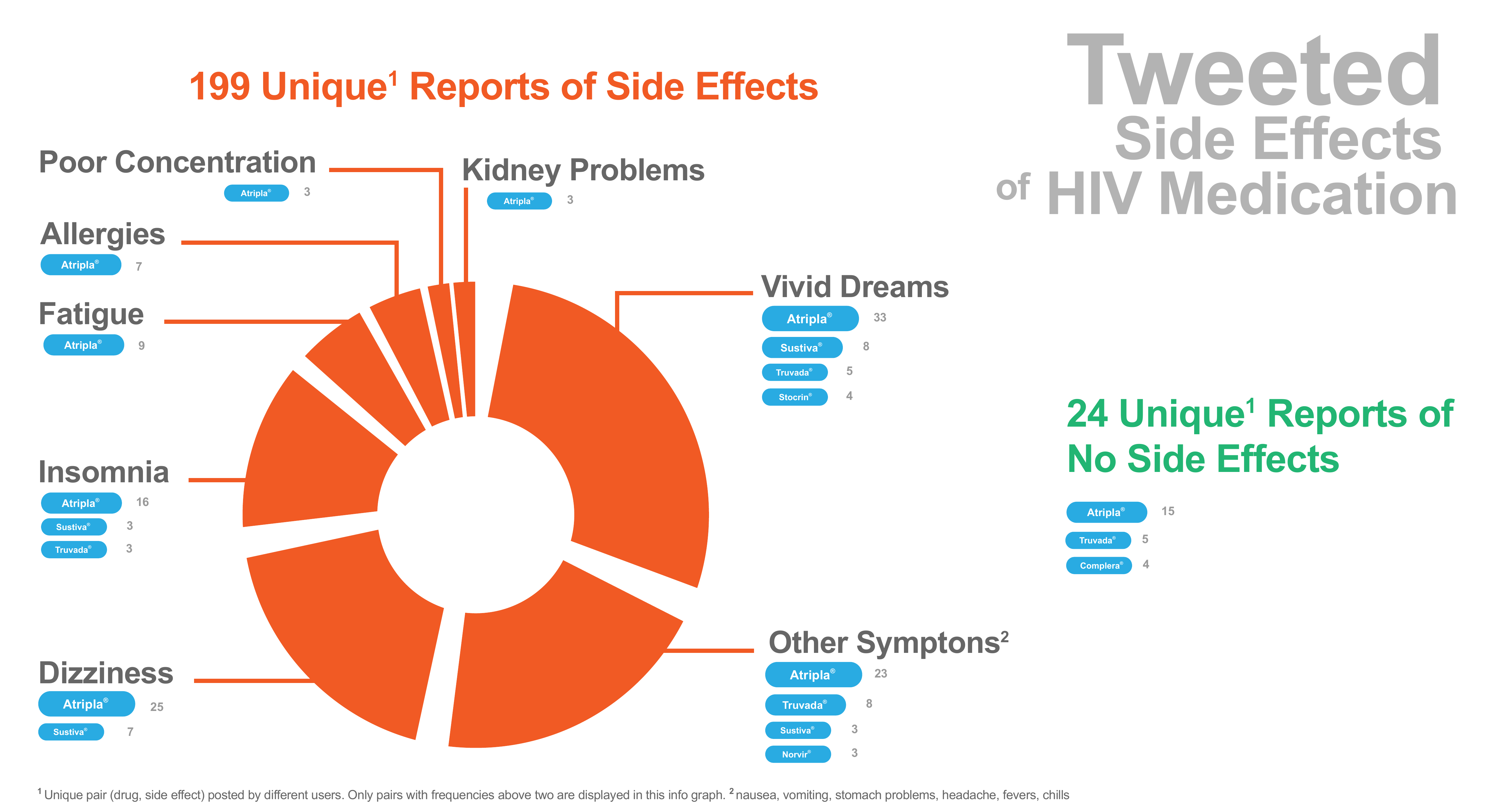}
\caption{Infographic of HIV drugs with associated side effects as reported by unique users on Twitter between 09/09/2010 and 08/23/2013.}
\label{drugsides}
\end{figure*}

In this section we present an infographic (Fig.~\ref{drugsides}) that summarizes the most common side effects associated to certain HIV drugs. We designed this infographic to be accessible to the mainstream community. To avoid double counting, a \emph{drug-effect} pair is considered only once for a given user. We only present pairs with at least three appearances in our samples. The numbers near the drug labels represent the number of times the drug is mentioned.

\vskip 2mm
In conclusion, most users report problems regarding their sleep, be it a nightmare or a vivid dream, or lack of sleep. They also report nausea, headaches, as well as symptoms comparable to the effect of illegal psychoactive drugs. Also, about 10$\%$ of the selected relevant tweets indicate no side effects.

\subsection{Analysis of user sentiment}
With our sample of isolated HIV patients, we aimed to study sentiment over time. For this study, we hand-rated tweets from -5 to 5, in steps of 1. The former indicates extremely negative feeling and the latter extremely positive.

\vskip 2mm
For each of the 60-day bins presented in the previous section, we computed the sum of the sentiments of all tweets:

\begin{equation}
\Psi \equiv \sum_{i=0}^{N\,\,tweet\,\,in\,\,bin} sentiment_{(tweet_i)} 
\end{equation}

We assign a systematic uncertainty of 1 to each rating, which leads to a total uncertainty in $\Psi$ of $\sqrt{N}$, where $N$ is the number of tweets in a given bin. Ideally, the systematic uncertainty would arise from the distributions of several ratings. This approach is out of the scope of this study in terms of precision and budget.

\begin{figure}
\centering
\includegraphics[width=3in,height=2in]{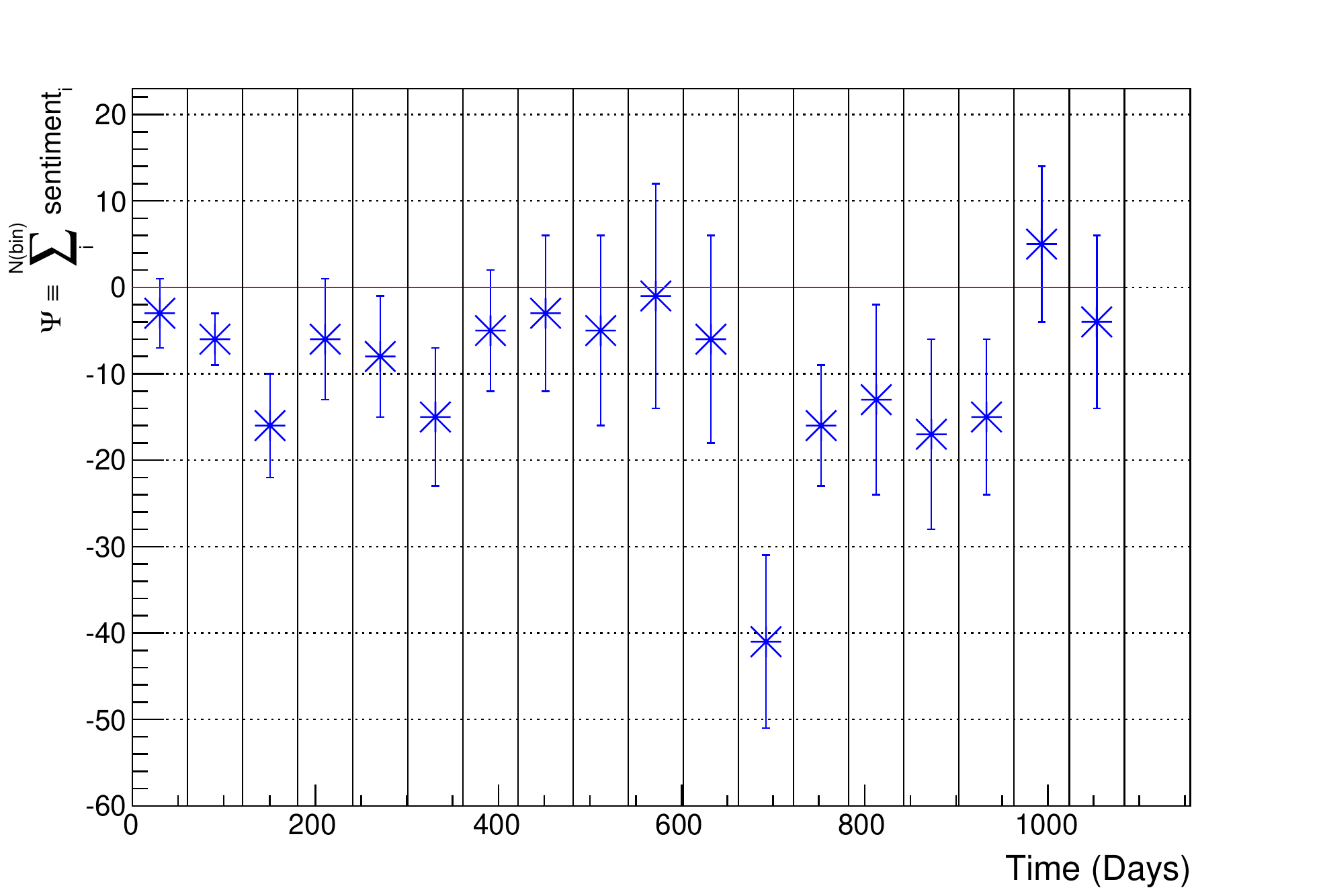}
\caption{Sentiment score $\Psi$ as a function of time. No requirements are applied to tweets.}
\label{sentimentall}
\end{figure}

\vskip 2mm
Fig.~\ref{sentimentall} shows the computed $\Psi$ in each bin defined as in Fig.~\ref{drugstime1} for tweets with no requirements imposed. This figure indicates an average negative sentiment, with departures from neutrality compatible within the uncertainties. This compatibility is less evident in bins 12 through 16. We study the effect on the sentiment of not considering retweets and conclude that the differences are minimal when compared to the total sample. Moreover, we compute the correlation between sentiment and number of drug appearances to be lower than 2$\%$. 

\begin{figure}
\centering
\includegraphics[height=2in]{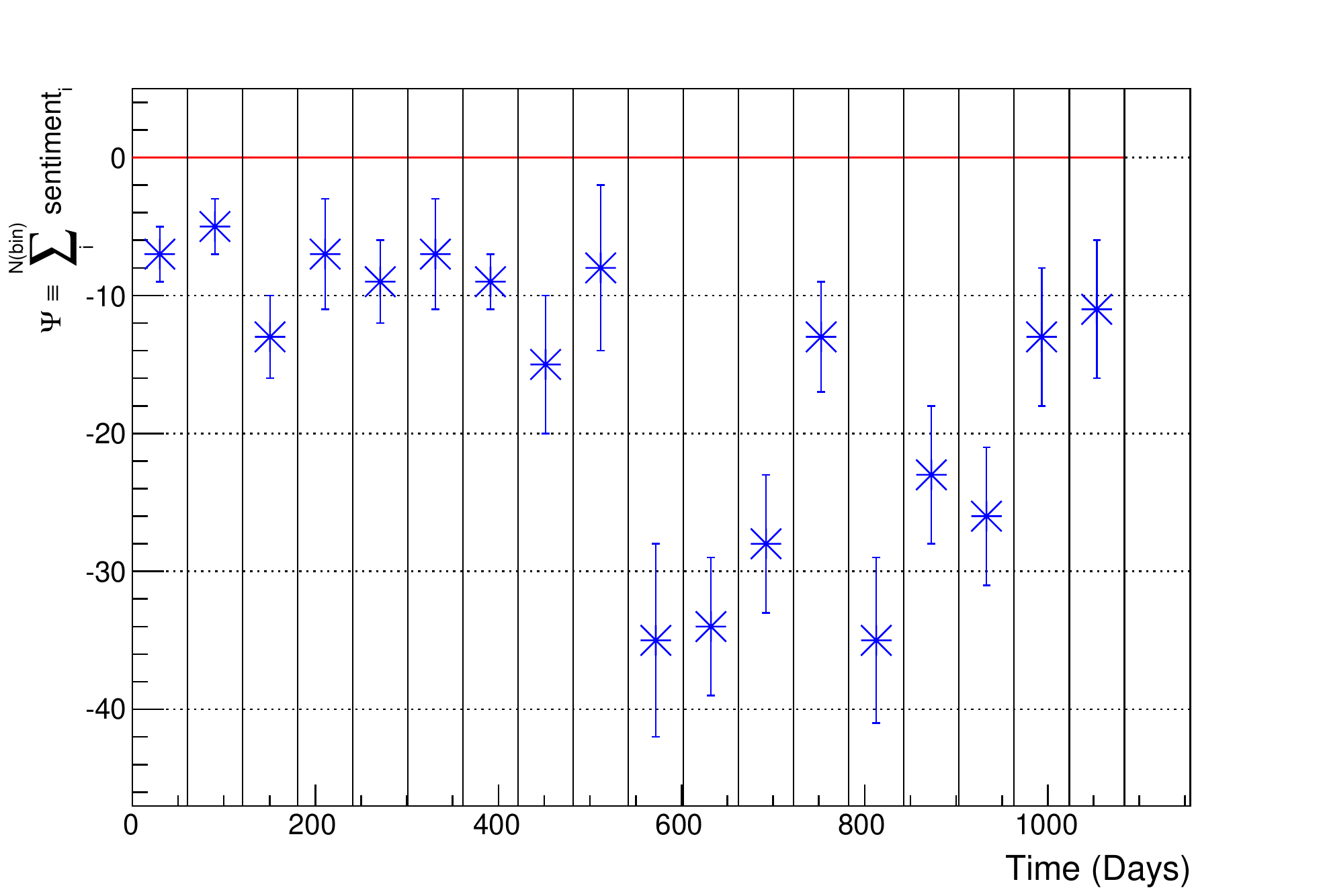}
\caption{Sentiment score $\Psi$ as a function of time. Retweets are excluded and only tweets with explicit side effects are considered.}
\label{sentimentallnortsides}
\end{figure}

\vskip 2mm
In Fig.~\ref{sentimentallnortsides} we exclude retweets and consider only tweets with explicitly mentioned side effects. The sentiment in this last figure clearly drifts towards negative values. Moreover, the dip in the distribution occurs at the same time as the peak of the distribution of mentions (Fig.~\ref{drugstimenortsides}).

\begin{figure}
\centering
\includegraphics[height=2in]{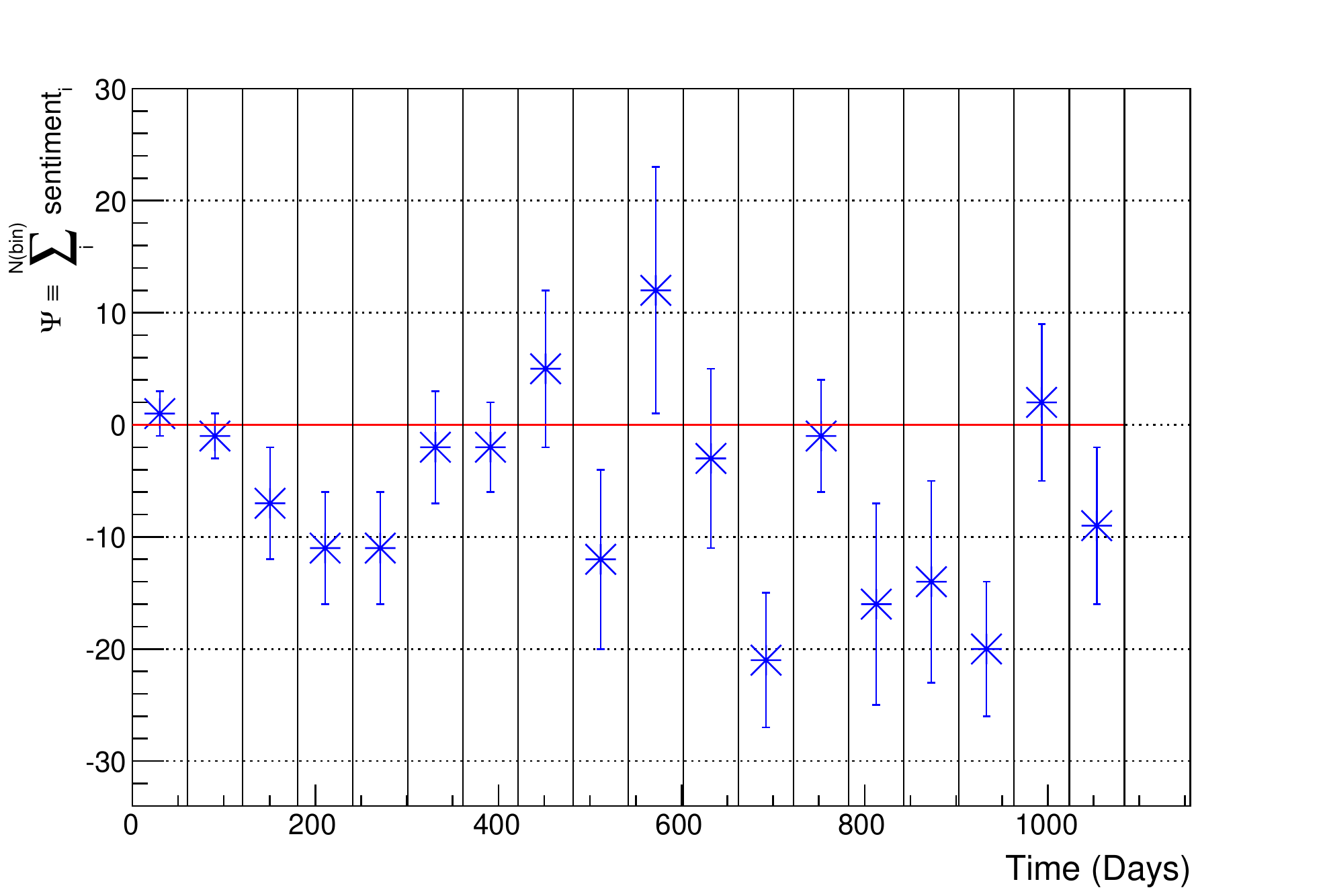}
\caption{Sentiment score $\Psi$ as a function of time for tweets with mentions of Atripla$\copyright$. No requirements are applied to tweets.}
\label{allatripla}
\end{figure}

\begin{figure}
\centering
\includegraphics[height=2in]{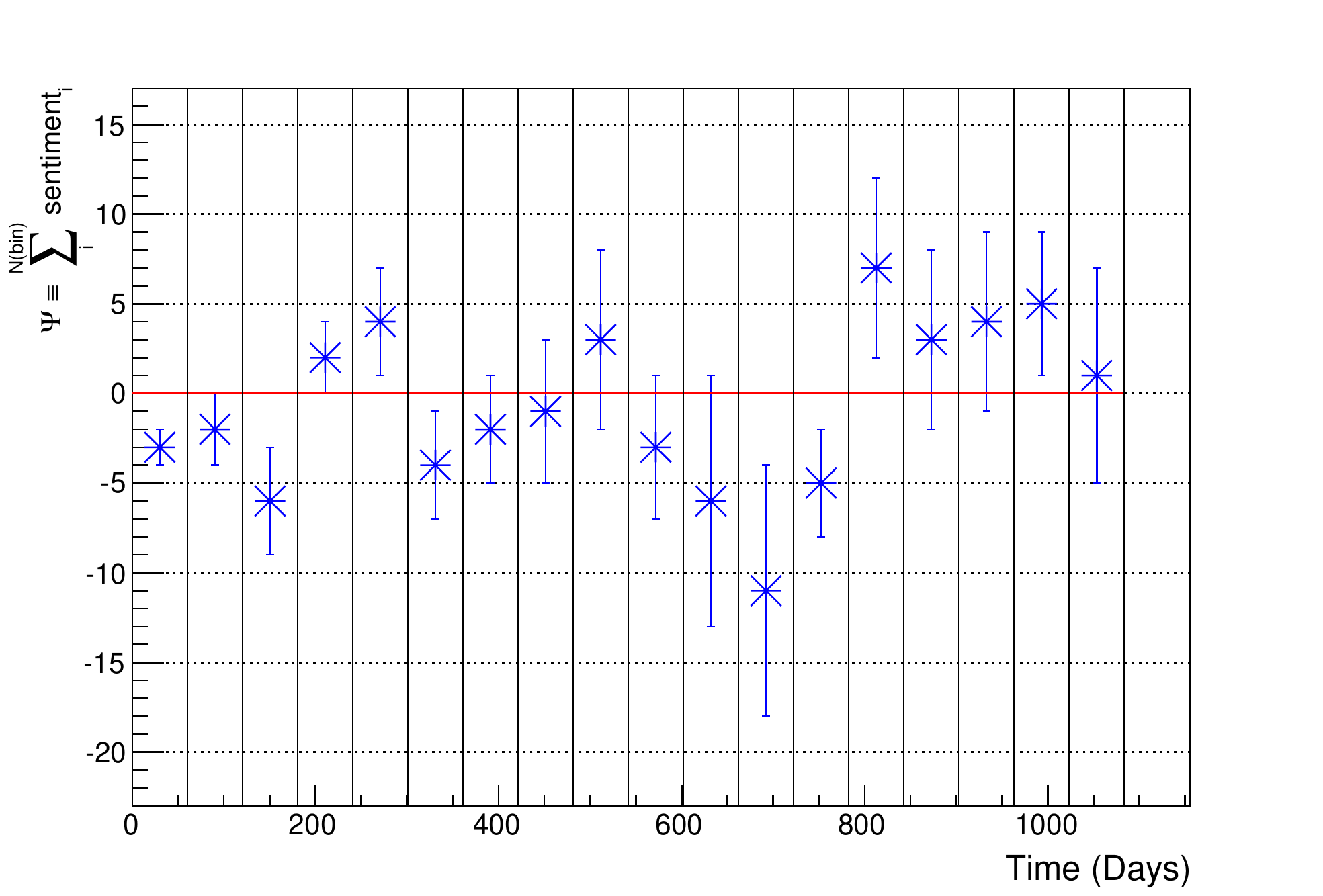}
\caption{Sentiment score $\Psi$ as a function of time for tweets with mentions of Truvada$\copyright$. No requirements are applied to tweets.}
\label{alltruvada}
\end{figure}

\begin{figure}
\centering
\includegraphics[height=2in]{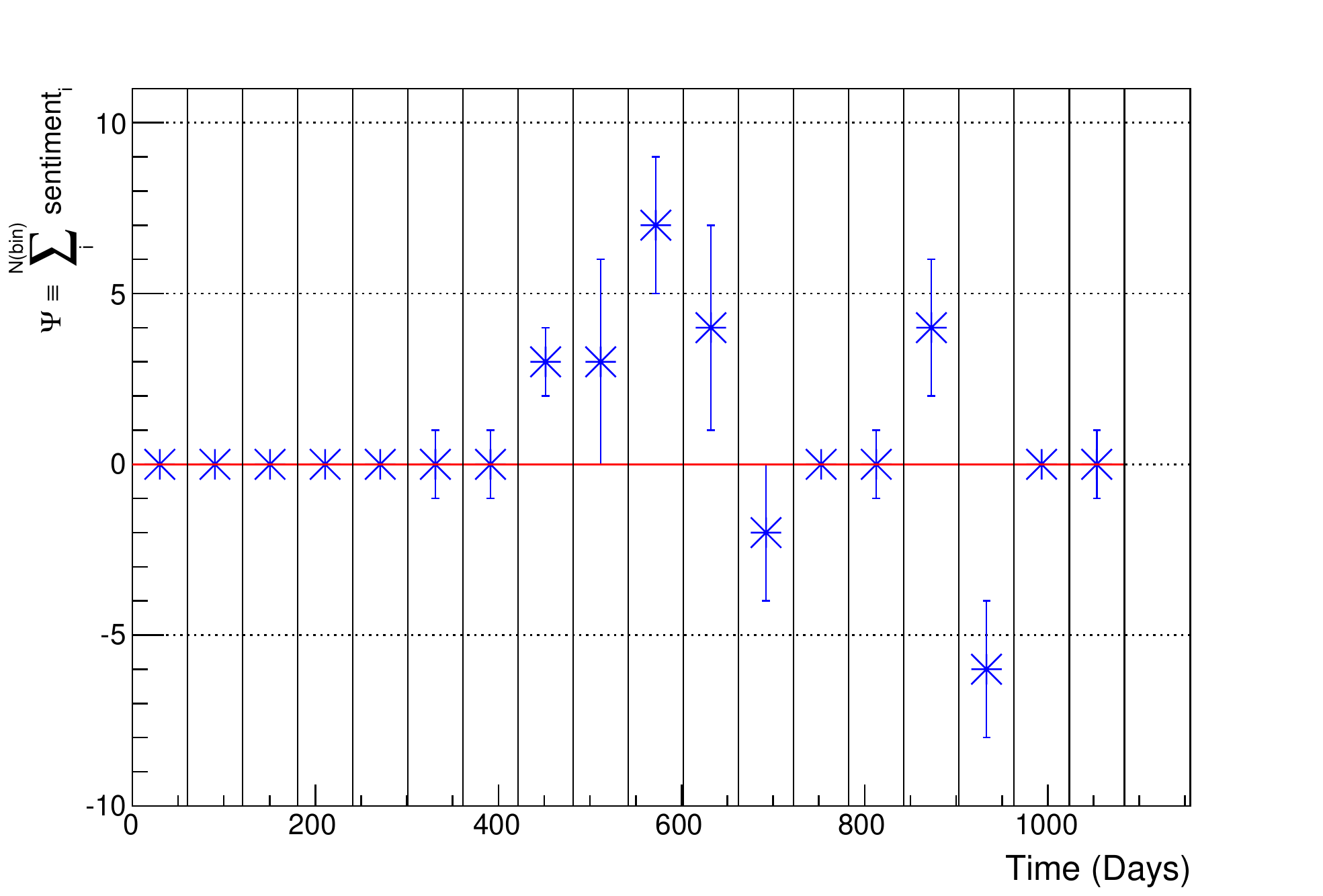}
\caption{Sentiment score $\Psi$ as a function of time with mentions of Complera$\copyright$. No requirements are applied to tweets.}
\label{allcomplera}
\end{figure}

\vskip 2mm
Plots for specific appearances of Atripla$\copyright$, Truvada$\copyright$, and Complera$\copyright$ indicate overall negative sentiment with the exception of Complera (see Figs~\ref{allatripla},~\ref{alltruvada},~\ref{allcomplera}).


\section{Conclusions}
We have presented a descriptive analysis of tweets collected from September 2010 through August 2013 and filtered by HIV-related keywords. After applying data cleaning, machine learning and human rating, we identify 1,642 tweets and 518 unique users describing personal information regarding HIV.

\vskip 2mm
This study explored ways in which to use Twitter data as a source of information for translational medicine. We presented an infographic summarizing the main side effects associated with HIV treatments. Our results conclude that the main concerns of twitter users with HIV medication are related to sleeping issues. Many users report symptoms such as nausea, fatigue, and dizziness at times likened to being under the effect of physcoactive drugs. 

\vskip 2mm
The sentiment associated to tweets of the identified targeted community shows no correlation with drug appearance. The overall underlying sentiment is negative and not compatible with neutrality within the uncertainties for tweets that contain references to side effects, and excluding retweets.
 
\section{Disclaimer}
The results presented in this paper should be taken as they are: a description of observations in data streamed from Twitter. In this paper, the authors report their findings and do not claim that the side effects quoted by Twitter users are directly caused by the listed drugs. Moreover, the authors declare no competing financial interests.

% ensure same length columns on last page (might need two sub-sequent latex runs)
\balance

%ACKNOWLEDGMENTS are optional
%\section{Acknowledgments}
%Marcel grant? Lausanne people? Zhoujie Huang for its input and help in hand-rating tweets. Isaac Bromley for designing Fig.~\ref{drugsides}.\\

%\bibliographystyle{abbrv}
%\bibliography{hiv_paper}  % vldb_sample.bib is the name of the Bibliography in this case

%\subsection{References}
%Generated by bibtex from your ~.bib file.  Run latex,
%then bibtex, then latex twice (to resolve references).

%APPENDIX is optional.
% ****************** APPENDIX **************************************
% Example of an appendix; typically would start on a new page
\pagebreak

\begin{appendix}

The dataset of tweets used in this paper was provided by Gnip and was size was 15\,$Gb$. We loaded several attributes of the tweet, and the tweet itself, into a PostgreSQL~\cite{postgresql} database. Tab.~3 shows the tweet attributes created in the database that are relevant for the study that we presented.

\begin{table}\label{varsdatabase}
\centering
\caption{Tweet attribute (left), attribute description (center) and type of variable defined in the database (right).}
\begin{tabular}{|l|l|l|}
\hline
 Name & Description &  Type \\
\hline
\hline
tweet & Posted tweet & text \\
user\_id & User's identification number & bigint \\
date & Day that tweet was posted & date \\
tweet\_id & Specific tweet identifier & bigint \\
user\_lang & Language on user's account & char varying(15) \\
\hline
\end{tabular}
 \end{table}

\section*{Processing of tweets}
We loaded each tweet into a database, and created an index to improve throughput query processing. We used the function GIN~\cite{postgresql} to index our database. Then we applied a tsvector~\cite{postgresql} function to each tweet. This function transformed each tweet into a list of unique lexemes. This allowed to check whether the keyword itself triggered the searching algorithm or was a compound word with one of the keywords embedded within. We requested at least one token to match one of our keywords. This requirement lead to a sample of 1.8 million. Then we discarded tweets that contained FTC as mentioned in Sec.~\ref{processing}, and estimated the most common tokens (Tab.~4).

\begin{table}\label{tokens}
\centering
\caption{Most common tokens (left column). Total number of appearances (right column) and number of rows or tweets it appears (middle column).}
\begin{tabular}{|l|l|l|}
\hline
    Token & Rows & Total \\
\hline
\hline
hiv & 199159 & 227374\\
t.co & 197078 & 202777\\
drug & 94680 & 104102\\
rt & 79181 & 85599\\
treatment & 69529 & 73126\\
truvada & 44325 & 49363\\
anti & 44591 & 45885\\
onlin & 27892 & 32538\\
buy & 27363 & 30218\\
anti-hiv & 29583 & 29900\\
prevent & 28018 & 29151\\
new & 26263 & 27667\\
fda & 25850 & 27620\\
approv & 24507 & 26773\\
retrovir & 21734 & 25647\\
aid & 23863 & 24804\\
bit.ly & 22462 & 22932\\
news & 16812 & 18168\\
generic & 14851 & 17865\\
de & 12959 & 17314\\
  \hline
\end{tabular}
\end{table}

\vskip 2mm
Hereafter signal denotes our targeted tweets posted by users with HIV, and noise denotes tweets about topics unrelated with our work, such as news. Our analysis started with loose criteria to reduce the noise in our sample. We selected three random samples of 500 tweets containing t.co, bit.ly, and starting by HIV. We selected also one random sample of 500 tweets not containing these words. We annotated all tweets in these random samples as either noise or signal. In the case of a slight indication of subjectivity, the tweet was rated as signal. We found 0, 1, and 1 possible signal tweets in the first samples, respectively. 24 possible signal tweets were found in the last sample where the words t.co, bit.ly and starting by HIV were excluded. From this last annotation we derived that 7.4 $\pm$ 2.7 possible signal tweets were expected to be found in 500 tweets taken from the original 316,081. The subsample with t.co, bit.ly, starting HIV was far from containing possible signal tweets, and was discarded from the following steps of the analysis. Although we expected to discard about 300 or 400 possible tweets, this cleaning process provided robustness to our procedure.

%\vskip 2mm 
%From this value of 24 we have about 5$\%$ (24/500). Therefore, one expects about 4,700 subjective tweets in subsample (94,178). If this is the number of subjective tweets in sample 316,081, then one would expect 7.4 $\pm$ 2.7 in a subsample of 500 tweets taken from 316,081.
%\vskip 2mm
%: $z$
%\vskip 2mm
%We discard the subsample t.co... We are discarding hence about 400 subjective tweets. But this are not our ultimate goal: HIV patients taking medication.

\vskip 2mm
In a second step, we checked tweets containing http, news, or buy. We selected three random samples of 140 tweets and annotated zero tweets in each of the three cases. As we found 24 possible signal tweets in a total of 500 tweets, we expected 6.7 $\pm$ 2.6 possible signal tweets in 140 tweets, assuming a Poissonian distribution. We computed the probability to discard a possible signal tweet if we removed tweets containing http, news or buy to be $4 \times 10^{-6}$:

\begin{equation}\label{eq:probabilities}
z > \frac{6.7-0}{2.6/\sqrt{3}} \rightarrow prob<0.000004,
\end{equation}

\vskip 2mm
were z is the cumulative probability distribution. After we discarded tweets containing either http, news, or buy, our dataset was reduced to 60,000 tweets.

\vskip 2mm
In the final step we separated two samples of 150 tweets containing either free, buy, de, e, za, que, en, lek, la, obat, da, majka, molim, hitno, mil, africa. We annotated zero possible signal tweets in both cases. Then, we estimated that the probability of losing possible signal tweets if we removed tweets containing at least one of this large set of words was less than $5 \times 10^{-6}$: 

\begin{equation}\label{eq:probabilities}
z > \frac{10.0-0}{3.2/\sqrt{2}} \rightarrow prob<0.000005
\end{equation}

\vskip 2mm
At the end of our cleaning process, 37,337 tweets remained within our dataset.

\pagebreak

\section*{Identification of targeted community}
After we applied the cleaning process described in the previous section, our dataset contained a larger fraction of possible signal tweets. The main goal of our analysis was to get a pure sample of these tweets. In this section we detail the steps taken to reach such sample.

\vskip 2mm
Firstly, we quantified each tweet into a set features, and, secondly, we combined these features into a single classifier.

\subsection*{Feature extraction}
Hereafter we give a broader definition of all features extracted from each tweet:

\begin{itemize}

\item \emph{modalcount}: number of times the words "should",\\"shoulda", "can", "could", "may", "might", "must", "ought", "shall", "would", and "woulda" occur in the tweet;
\item \emph{futurecount}: number of times the words "going", "will", "gonna", "should", "shoulda", "ll", "d" occur in the tweet;
\item \emph{personalcount}: number of times the words "i", "me", "my", "mine",  "ill", "im", "id", "myself" occur in the tweet; 
\item \emph{negative}: number of times the words "not", "wont", "nt", "shouldnt", "couldnt" occur in the tweet;
\item \emph{secondpron}: number of times the words "you", "youll", "yours", "yourself" occur in the tweet;
\item \emph{thirdpron}: number of times the words "he", "she", "it", "his", "her", "its", "himself", "him", "herself", "itself", "they", "their", "them", "themselves" occur in the tweet;
\item \emph{relatpron}: number of times the words "that", "which", "who", "whose", "whichever", "whoever", "whoever" occur in the tweet;
\item \emph{dempron}: number of times the words "this", "these", "that", "those" occur in the tweet;
\item \emph{indpron}: number of times the words "anybody", "anyone", "anything", "each", "either", "everyone", "everything", "neither", "nobody", "somebody", "something", "both", "few", "many", "several", "all", "any", "most", "none", "some" occur in the tweet;
\item \emph{intpron}: number of times the words "what", "who", "which", "whom", "whose" occur in the tweet; 
\item \emph{percent}: number of $\%$ symbols in the tweet;
\item \emph{posnoise}: number of times the words "new", "pill", "state", "states", "stats", "drug", "people", "approved", "approve", "approves", "approval", "approach", "prevention", "prevent", "prevents", "prevented" occur in the tweet;
\item \emph{is\_notenglish}: number of times words contained in a list of words extracted from annotated tweets as not English occur in the tweet;
\item \emph{regularpast}: number of words ending with $ed$ contained in the tweet;
\item \emph{gerund}: number of words ending with $ing$ contained in the tweet;
\item \emph{nment}: number of words ending with $ment$ contained in the tweet;
\item \emph{nfull}: number of words ending with $full$ contained in the tweet;
\item \emph{tagadj}: ratio of the number of adjectives tagged using NLTK~\cite{nltk} in the tweet by the total number of words in the tweet;
\item \emph{tagverb}: ratio of the number of verbs tagged using NLTK~\cite{nltk} in the tweet by the total number of words in the tweet;
\item \emph{tagprep}: ratio of the number of prepositions tagged using NLTK~\cite{nltk} in the tweet by the total number of words in the tweet;
\item \emph{tagnoun}: ratio of the number of nouns tagged using NLTK~\cite{nltk} in the tweet by the total number of words in the tweet;
\item \emph{tagconj}: ratio of the number of conjunctions tagged using NLTK~\cite{nltk} in the tweet by the total number of words in the tweet;
\item \emph{tagadv}: ratio of the number of adverbs tagged using NLTK~\cite{nltk} in the tweet by the total number of words in the tweet;
\item \emph{tagto}: ratio of the number of $to$ tagged using NLTK~\cite{nltk} in the tweet by the total number of words in the tweet;
\item \emph{tagdeterm}: ratio of the number of determinants tagged using NLTK~\cite{nltk} in the tweet by the total number of words in the tweet;
\item \emph{sis\_noise}: ratio of the similarity of the tweet with a corpus of annotated noise tweets by its uncertainty. To compute the similarity we first create a sparsity matrix of the tokens in the annotated corpus, then count the number of times the token appears in the tweet and divide by the number of elements in the corpus. We use scikit-learn~\cite{scikitlearn} library in several parts of the definition of sis\_noise; 
\item \emph{sis\_signal}: ratio of the similarity of the tweet with a corpus of annotated noise tweets by its uncertainty. To compute the similarity we first create a sparsity matrix of the tokens in the annotated corpus, then count the number of times the token appears in the tweet and divide by the number of elements in the corpus. We use scikit-learn~\cite{scikitlearn} library in several parts of the definition of sis\_noise;
\item \emph{is\_english}: number of words in corpus of English words\\~\cite{nltk} divided by number of words in corpuses of Spanish, Portuguese, French, German, Dutch, Italian, Russian, Swedish, and Danish~\cite{nltk}. We add one in both numerator and denominator to avoid dividing by zero.
\item \emph{bigrams\_noise}: number of bigrams found in tweet that are contained in list of bigrams of noise annotated bigrams corpuses divided by the total number of bigrams from annotated corpuses;
\item \emph{bigrams\_signal}: number of bigrams found in tweet that are contained in list of bigrams of signal annotated bigrams corpuses divided by the total number of bigrams from annotated corpuses;
\item \emph{isolation}: number of keywords contained in tweet minus one;
\item \emph{common\_noise}: sum of the weights of each word contained in most common 25\% of words in noise annotated tweets;  
\item \emph{common\_signal}: sum of the weights of each word contained in most common 25\% of words in signal annotated tweets;
\item \emph{wordscount}: number of words in tweet;
\item \emph{tweetlength}: number of characters in tweet.

\end{itemize}

We used crowdsourcing Amazon Mechanical Turk to rate tweets into three categories: signal, noise, and not English. Two workers rated each tweet. We had an agreement of about 80\% between workers that rated our tweets. We used tweets with agreement to define our samples. Moreover, we used tweets rated as not English as a control sample in order to remove foreign language tweets (see below).

We show all the variables in Fig.~\ref{hiv:variables:1}. These figures have been obtained with 109 signal events and 1,809 noise events. Several features show more separation power than others. For instance, sis\_noise and commonsignal falls in the former category, while nful and tagadj falls in the latter. Tab.~5 shows the computed separation power between signal and noise of each feature.

\begin{figure*}
\centering
\includegraphics[width=1.5in]{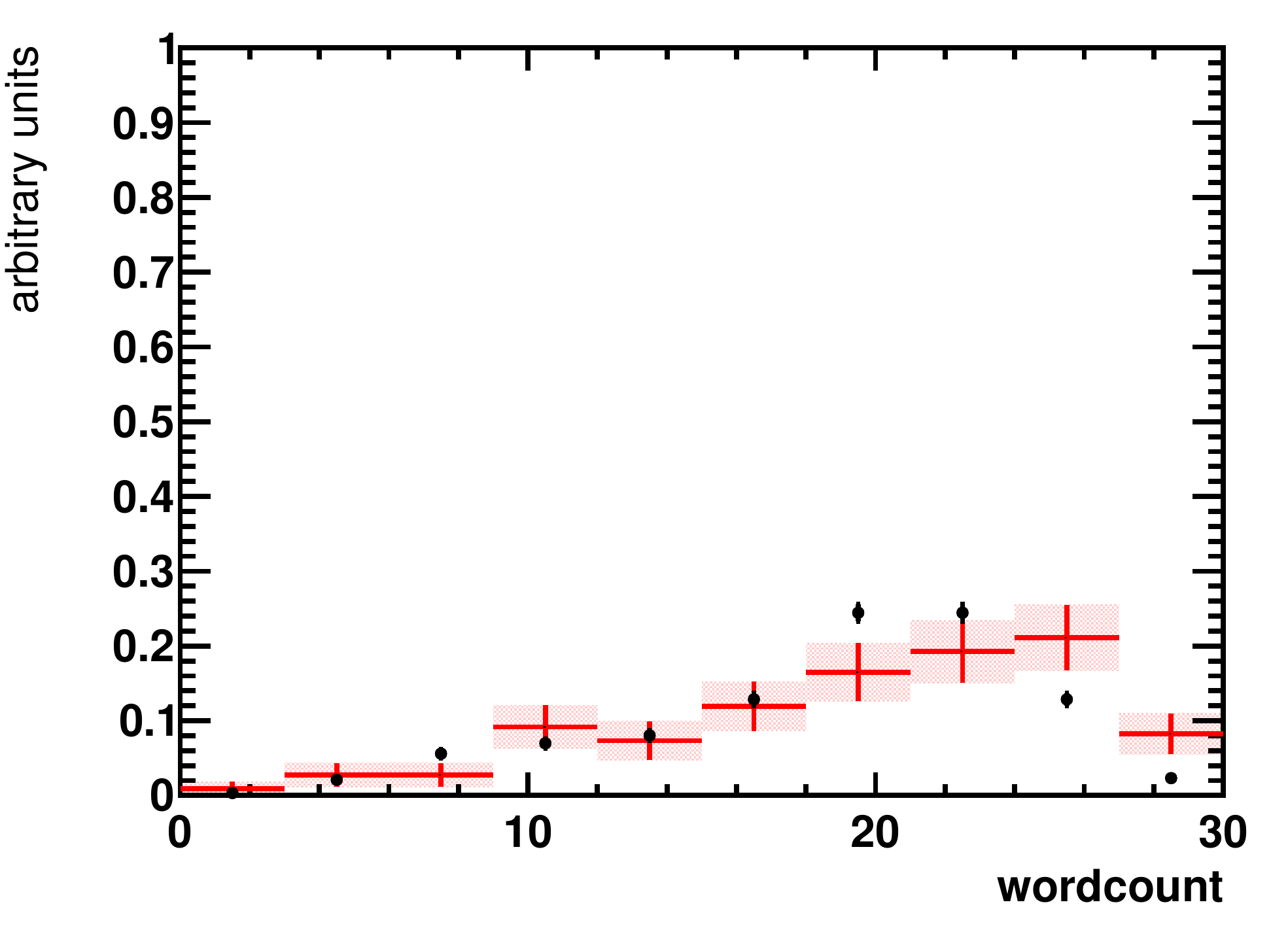}
\includegraphics[width=1.5in]{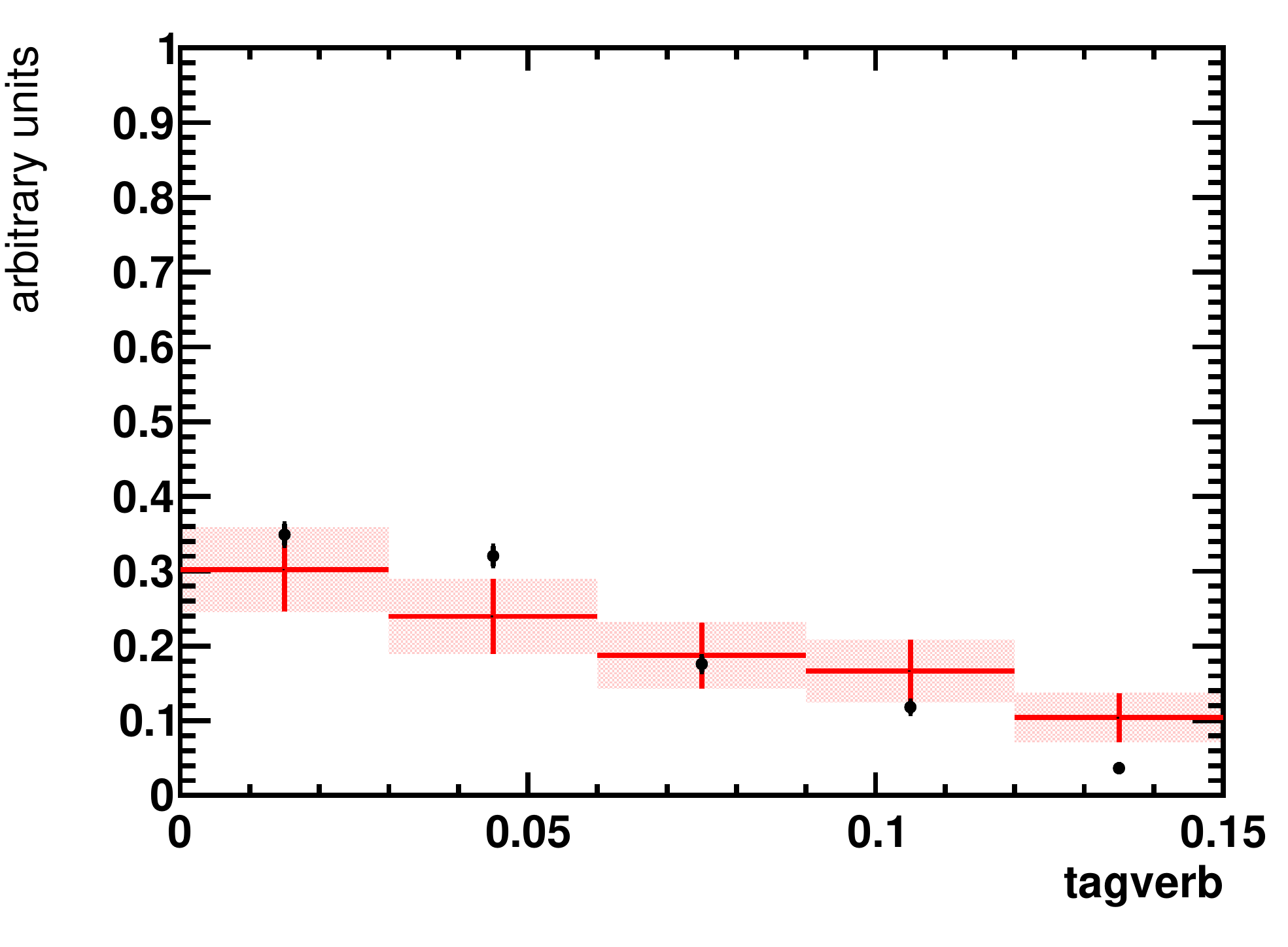}
\includegraphics[width=1.5in]{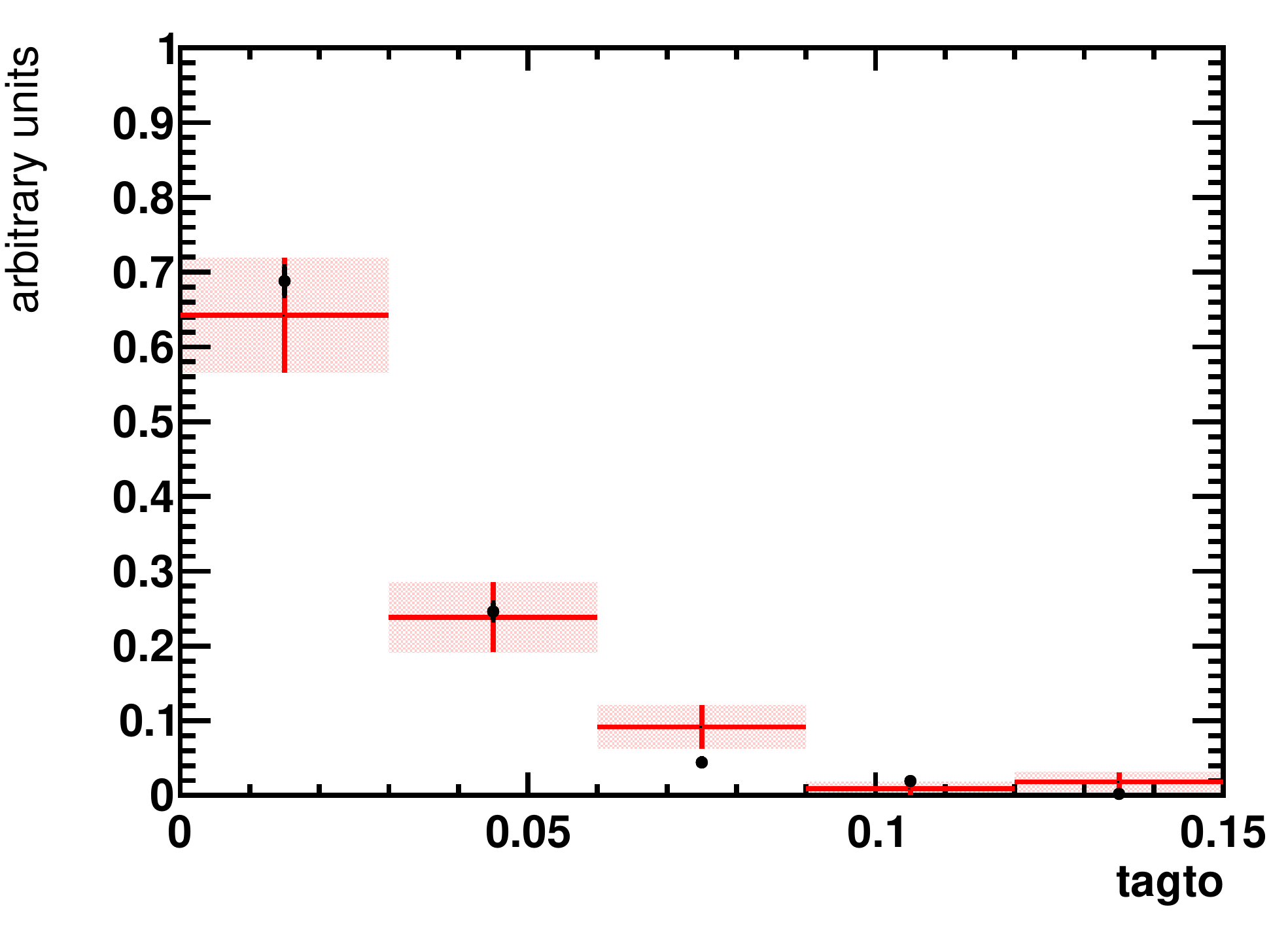}
\includegraphics[width=1.5in]{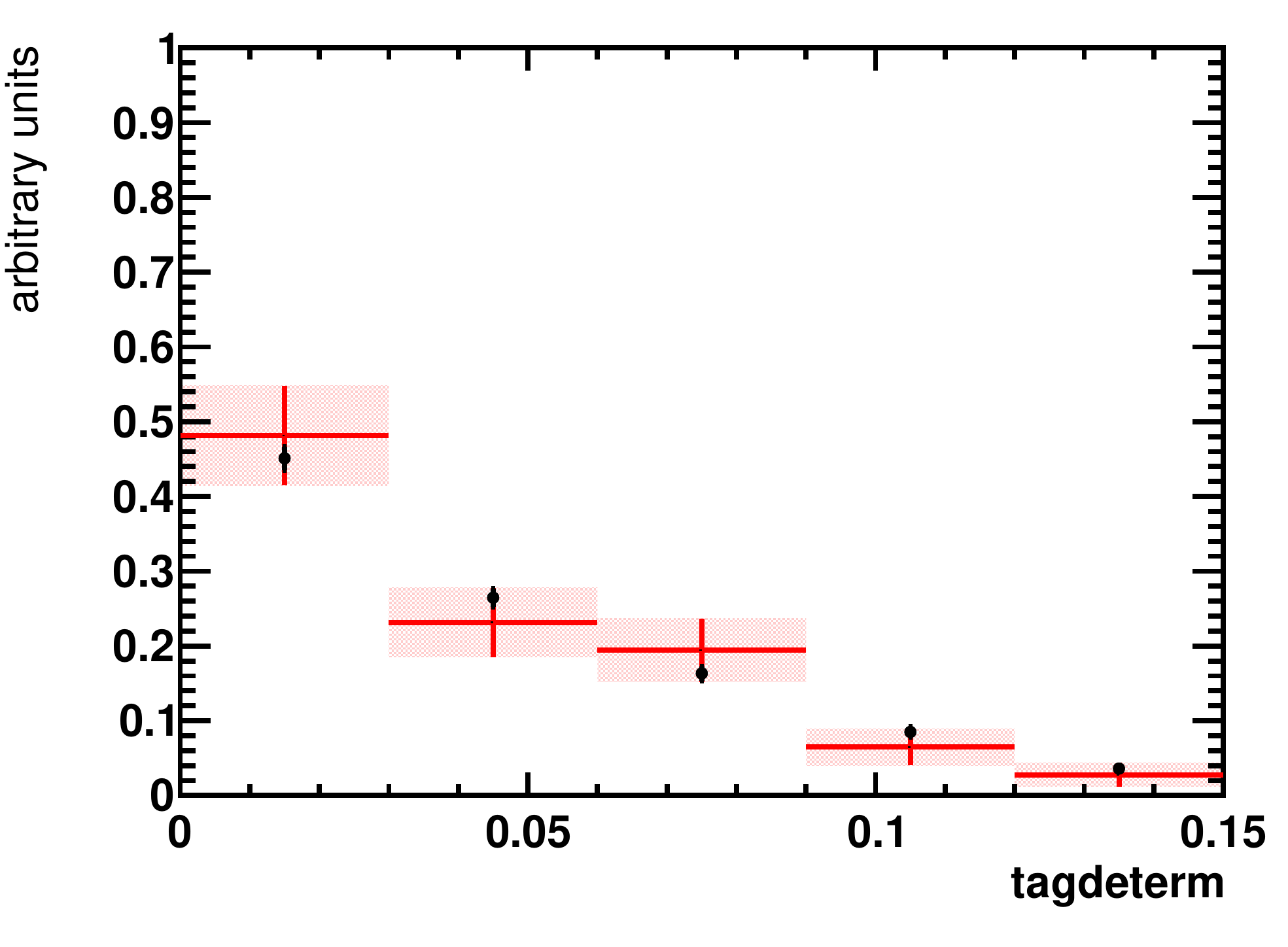}
\includegraphics[width=1.5in]{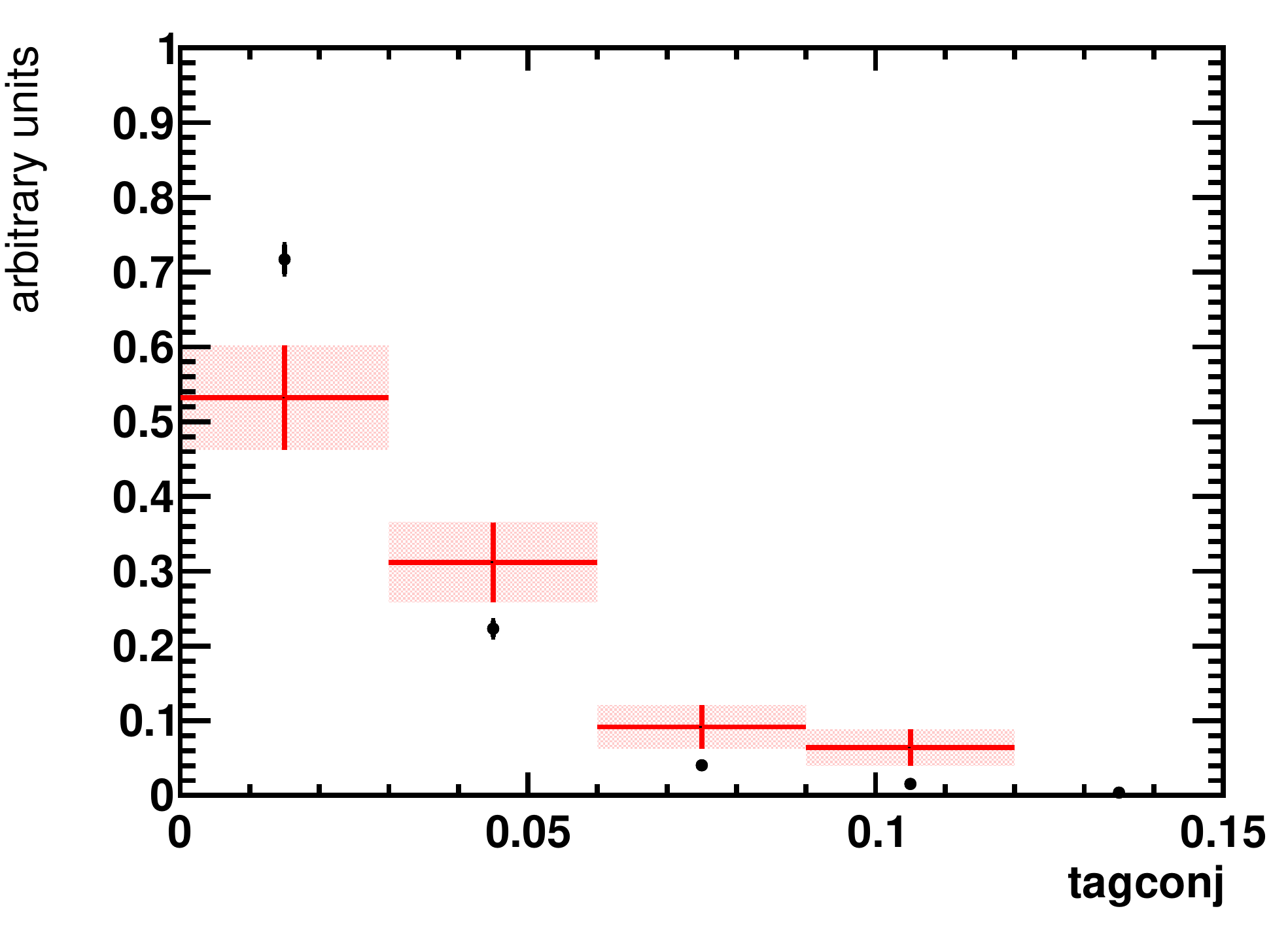}
\includegraphics[width=1.5in]{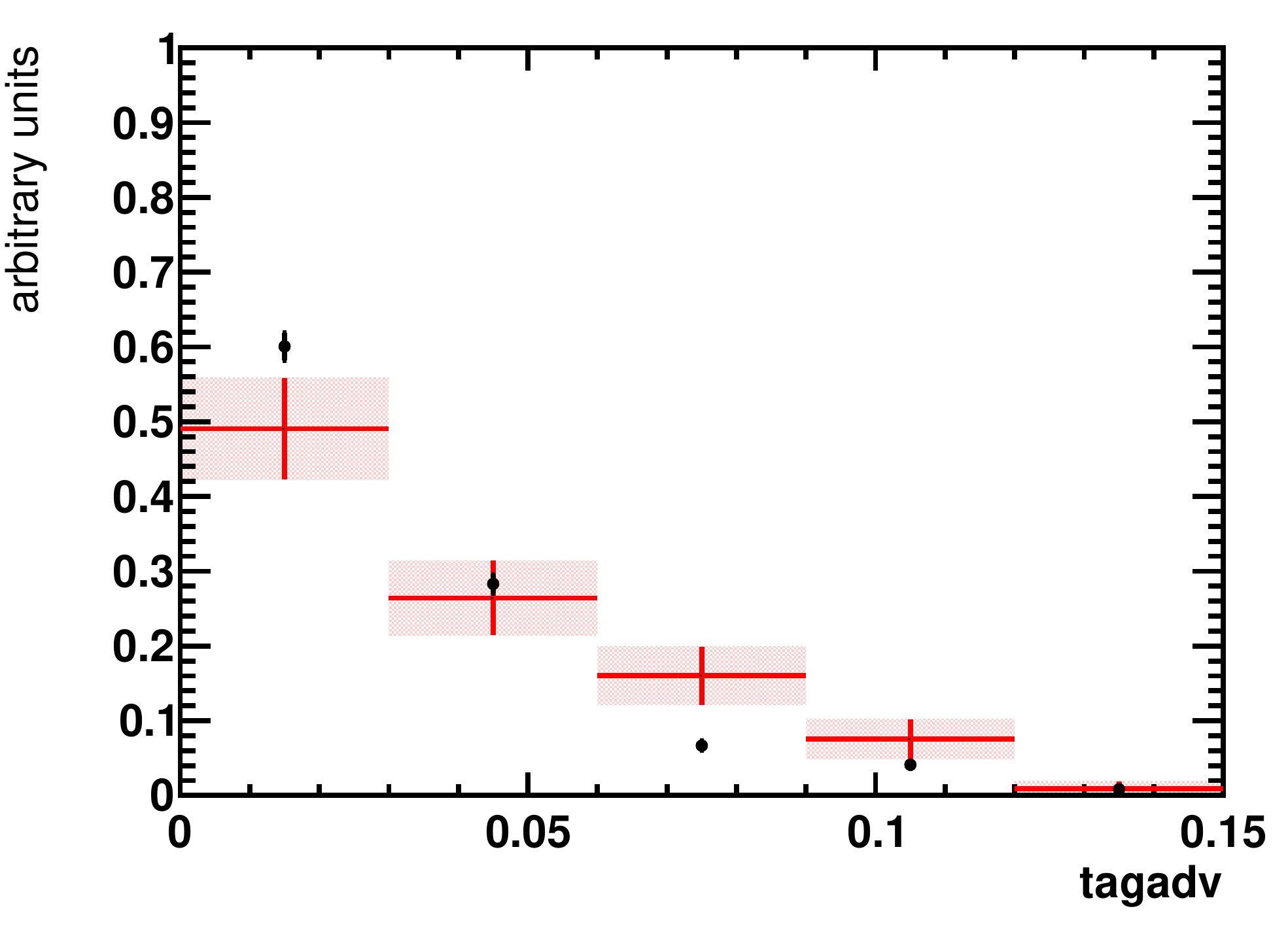}
\includegraphics[width=1.5in]{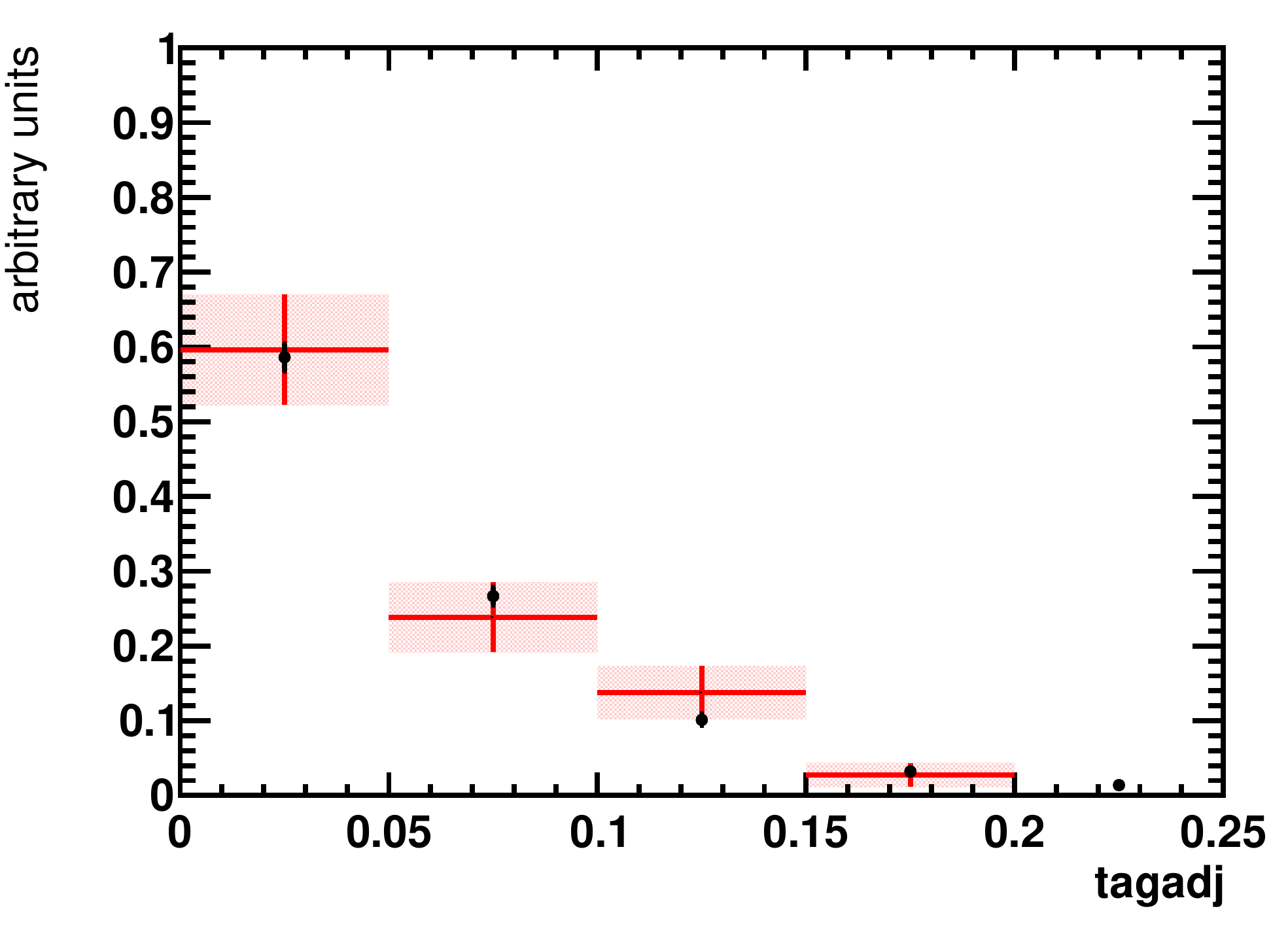}
\includegraphics[width=1.5in]{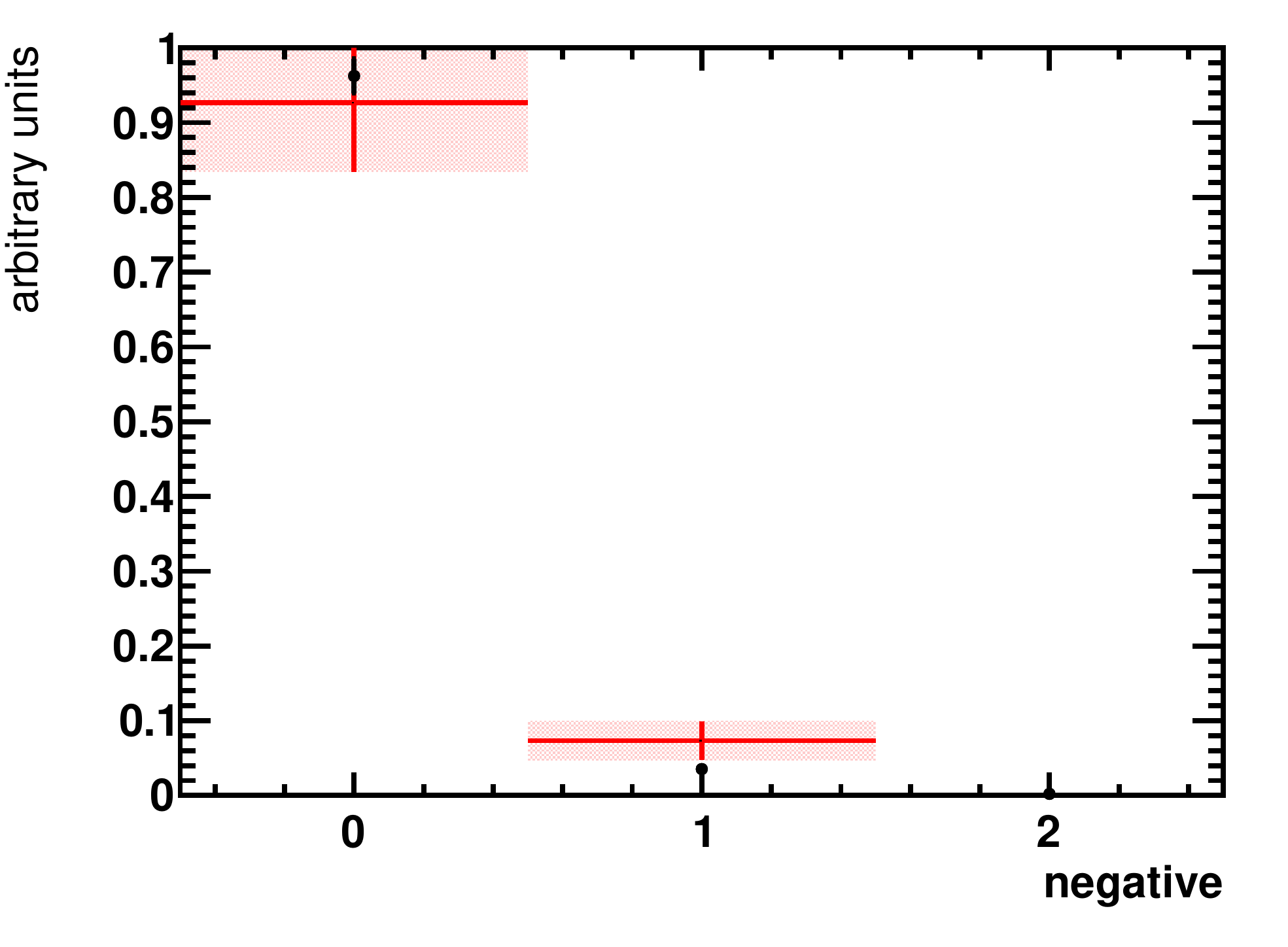}
\includegraphics[width=1.5in]{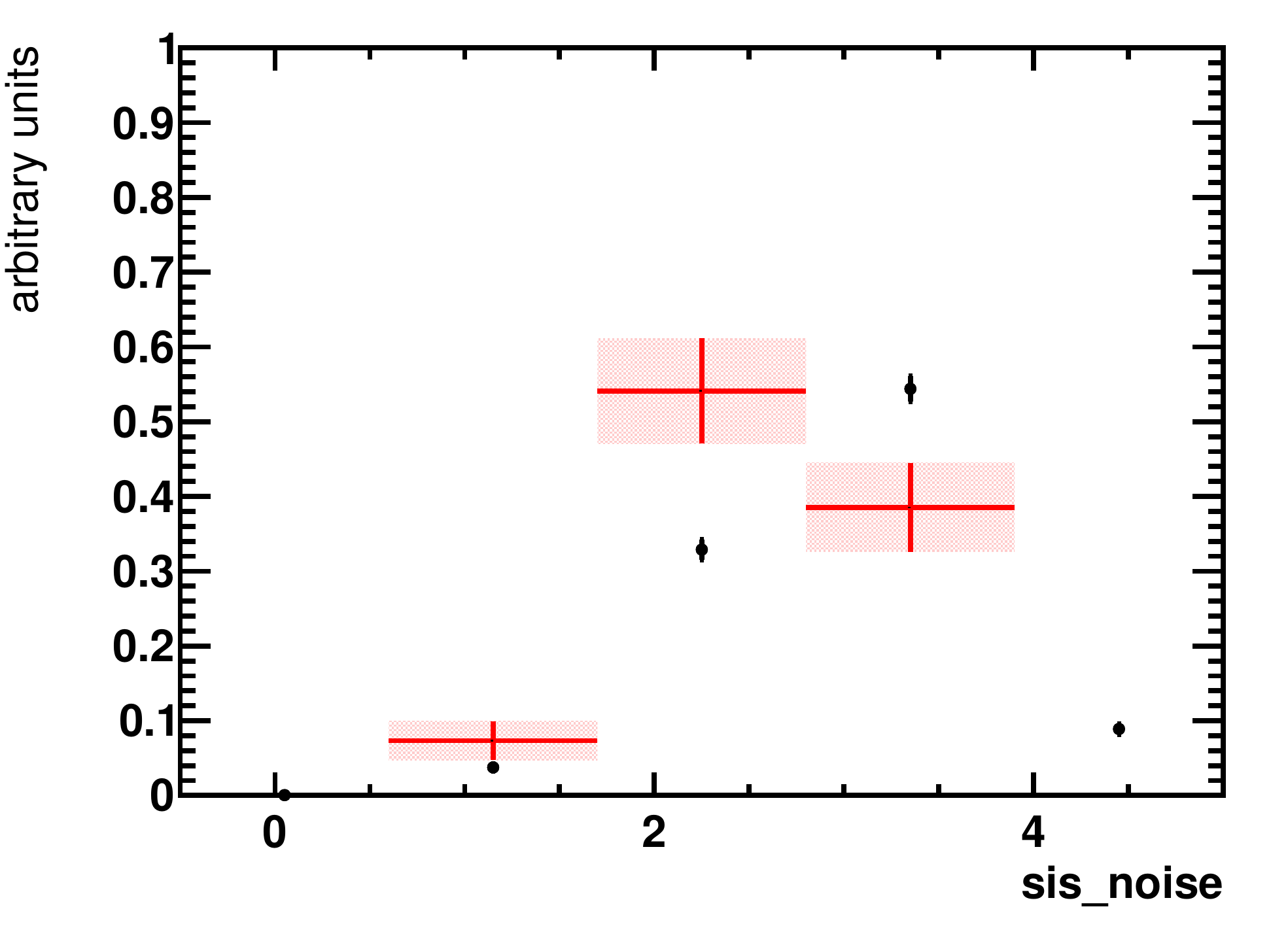}
\includegraphics[width=1.5in]{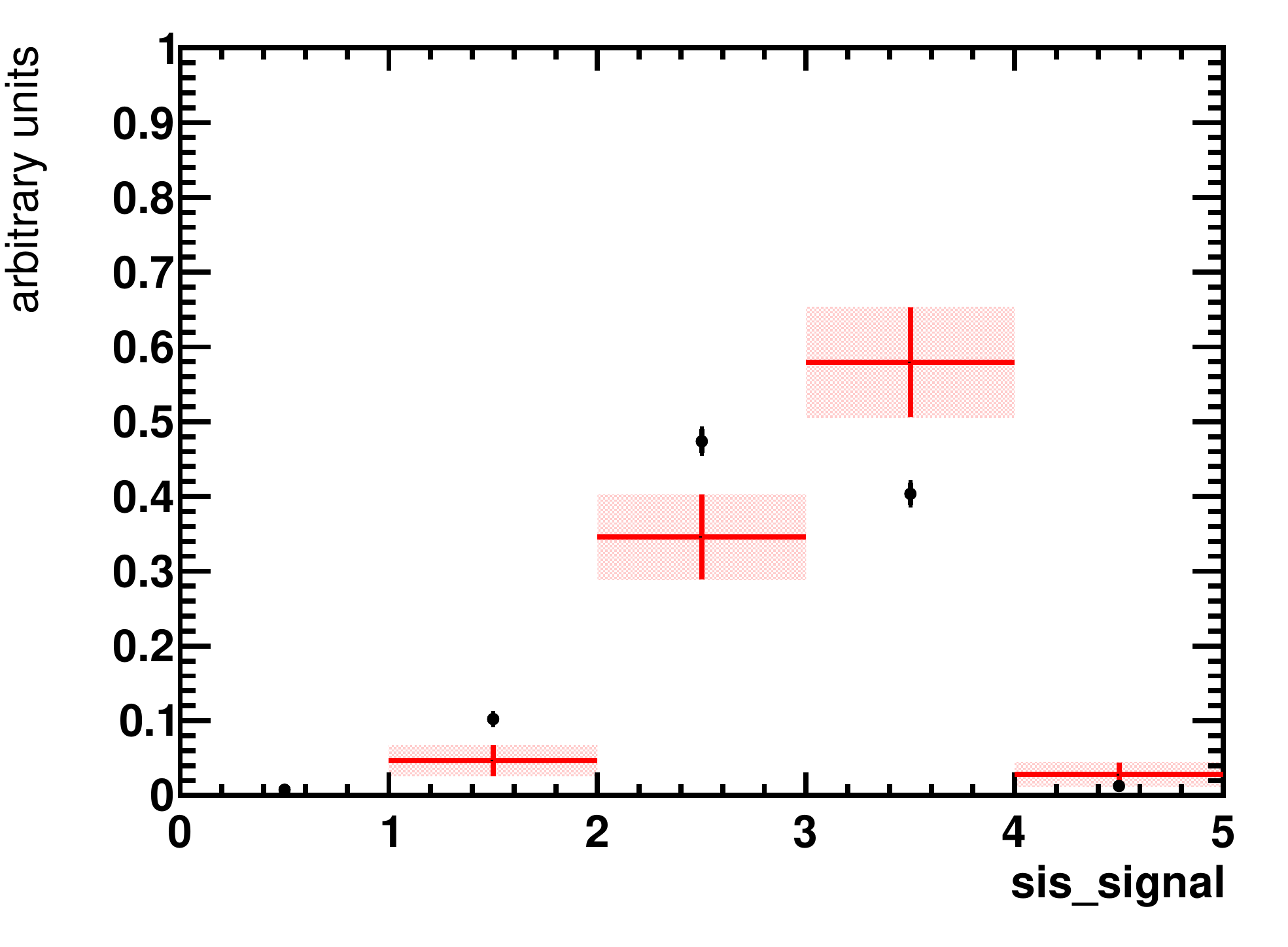}
\includegraphics[width=1.5in]{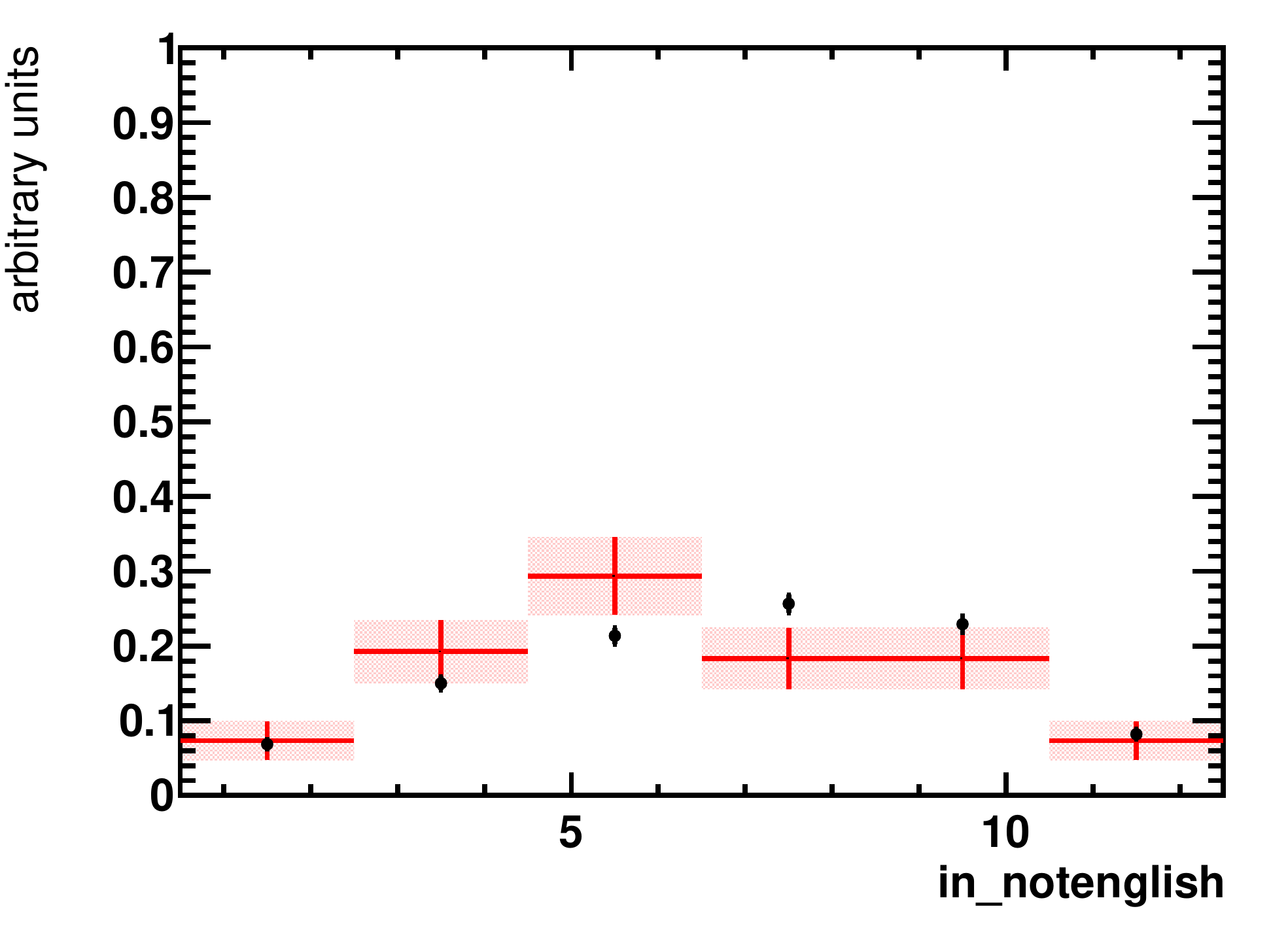}
\includegraphics[width=1.5in]{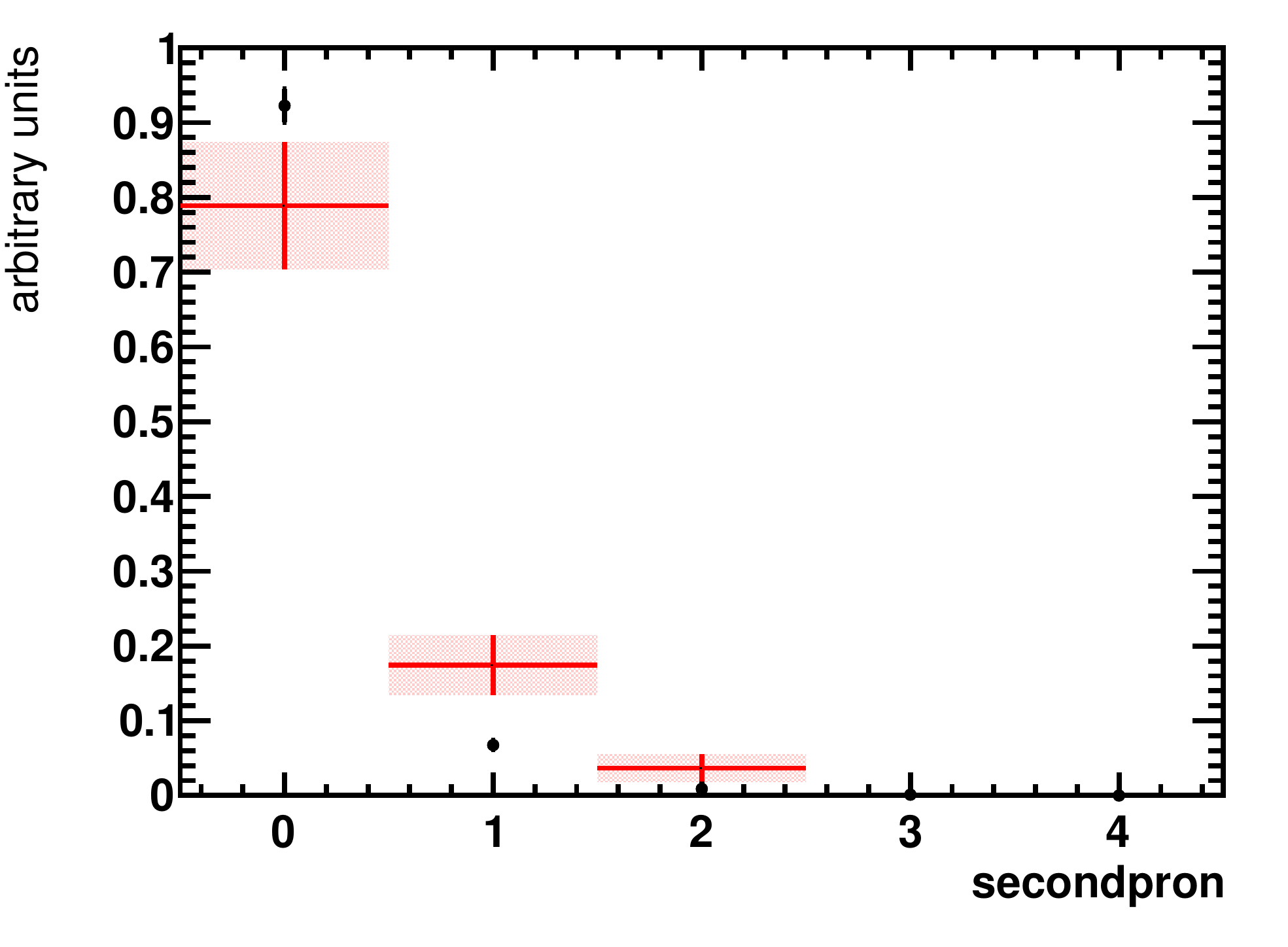}
\includegraphics[width=1.5in]{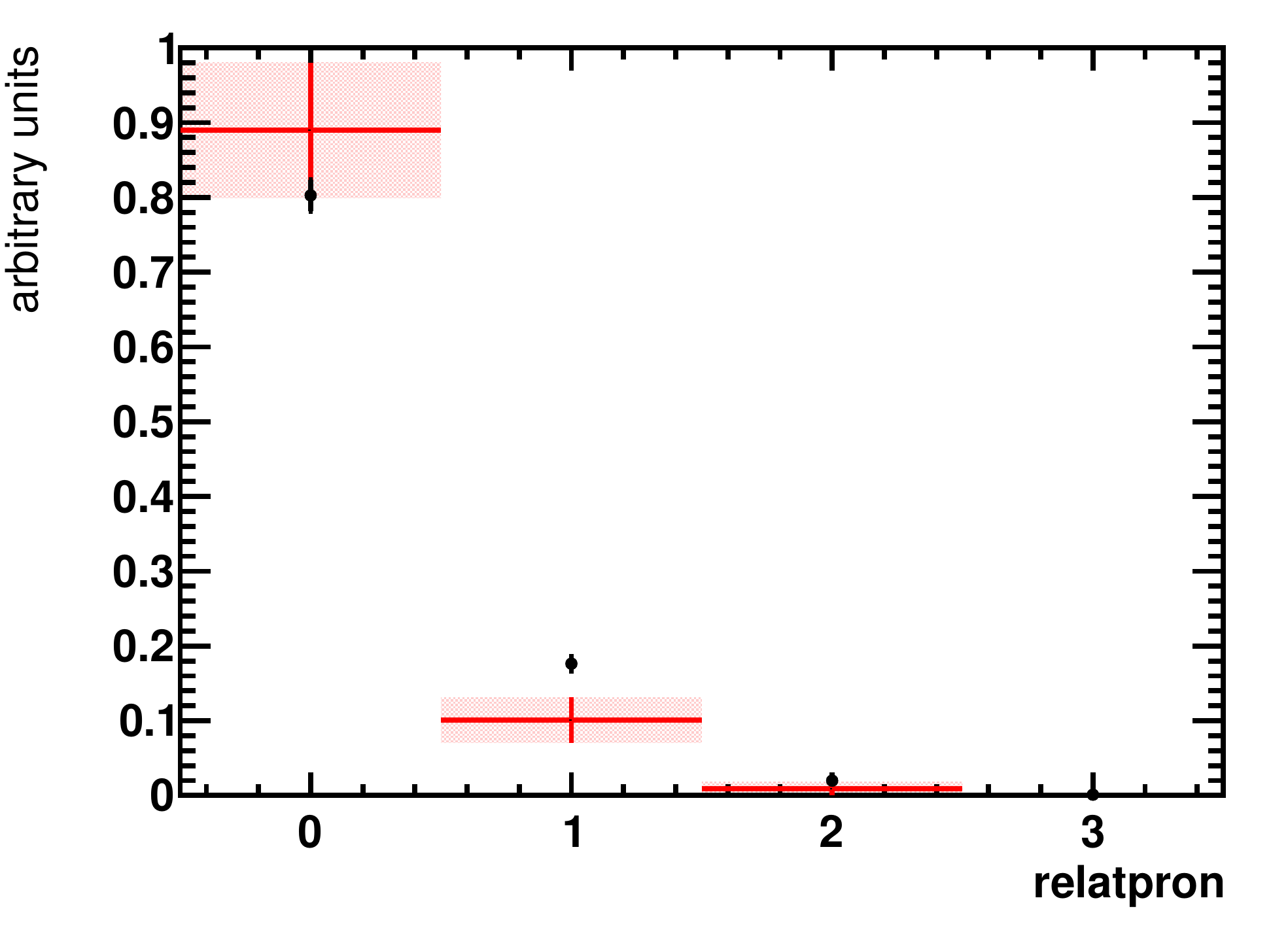}
\includegraphics[width=1.5in]{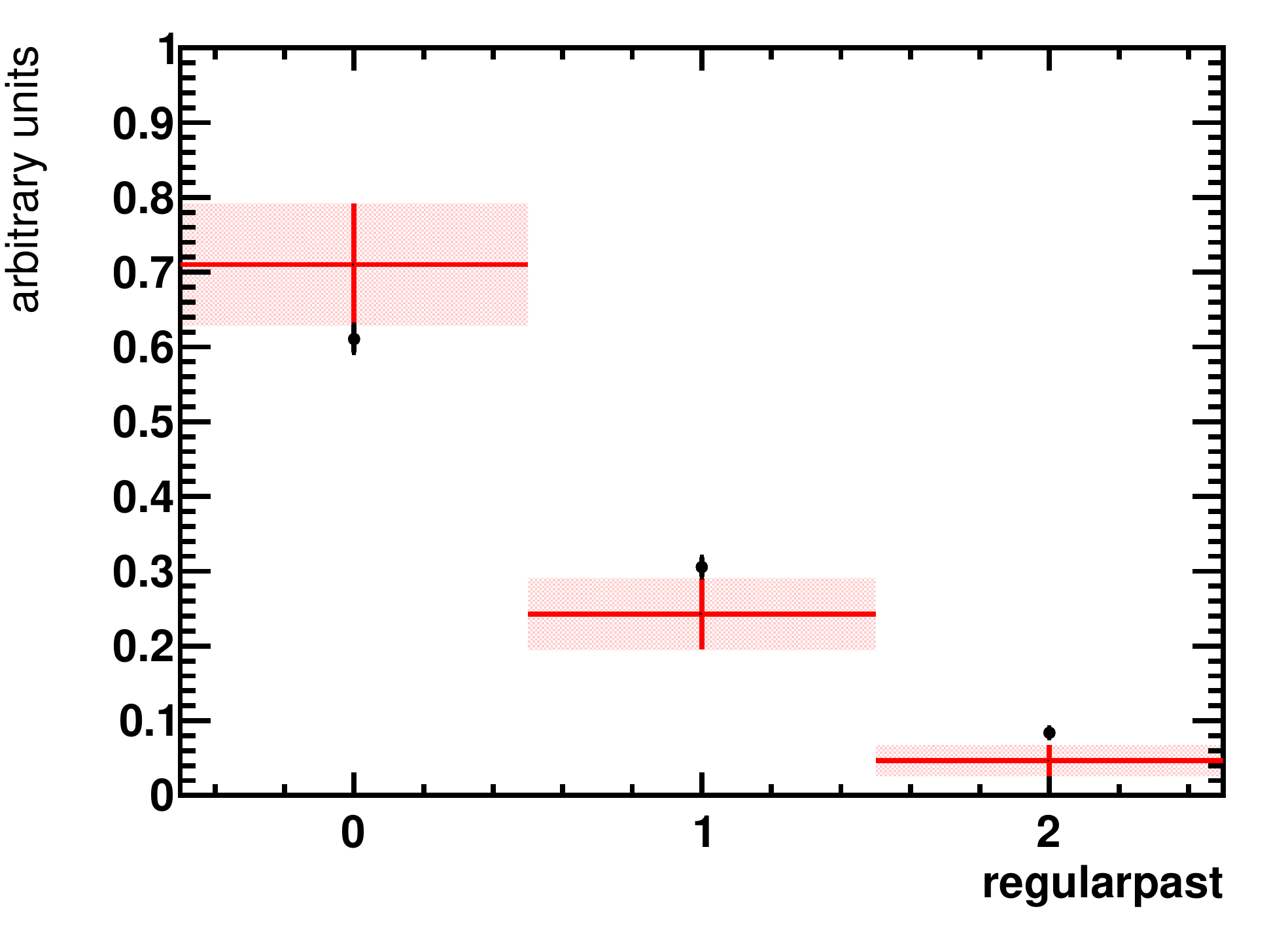}
\includegraphics[width=1.5in]{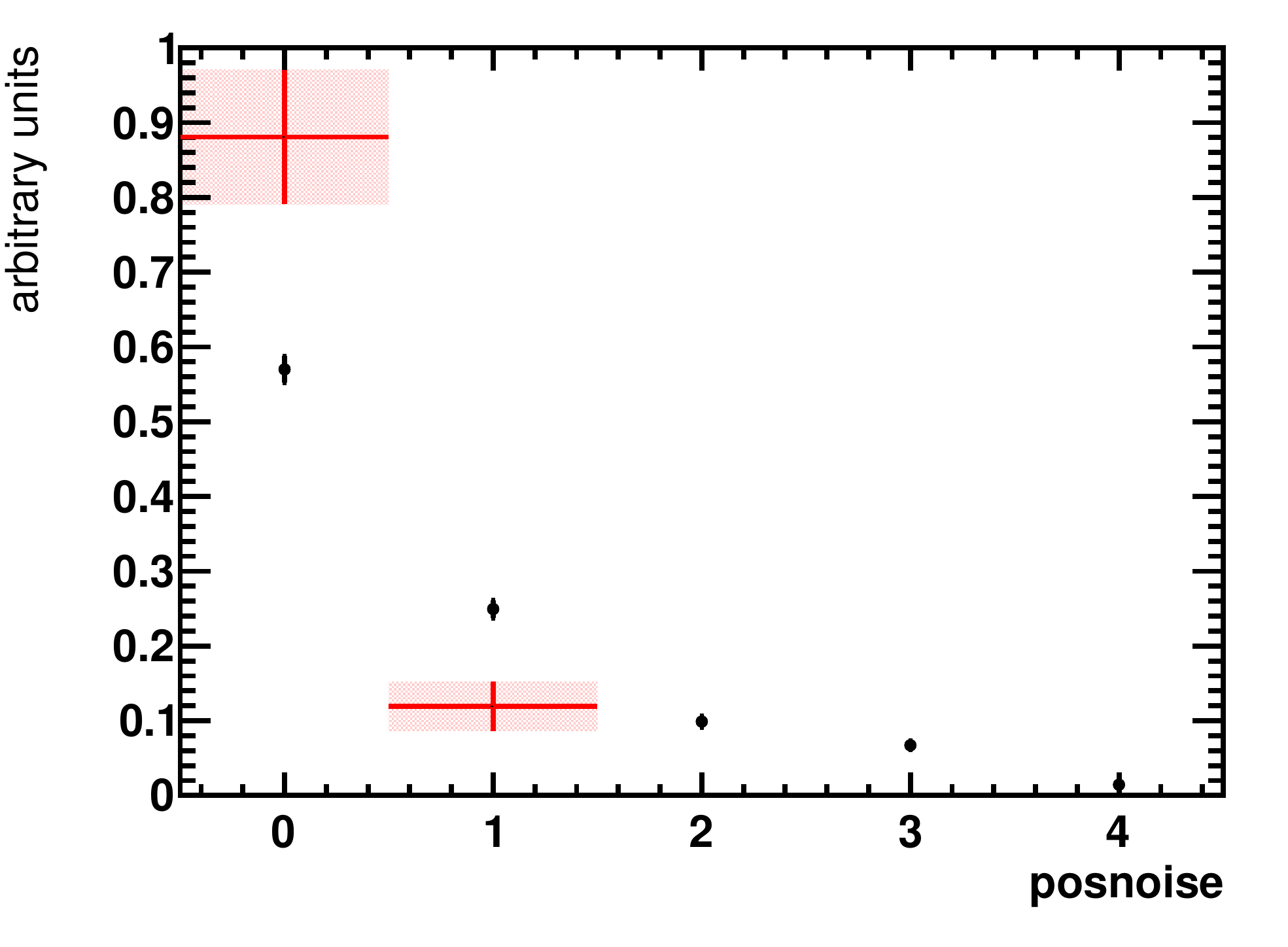}
\includegraphics[width=1.5in]{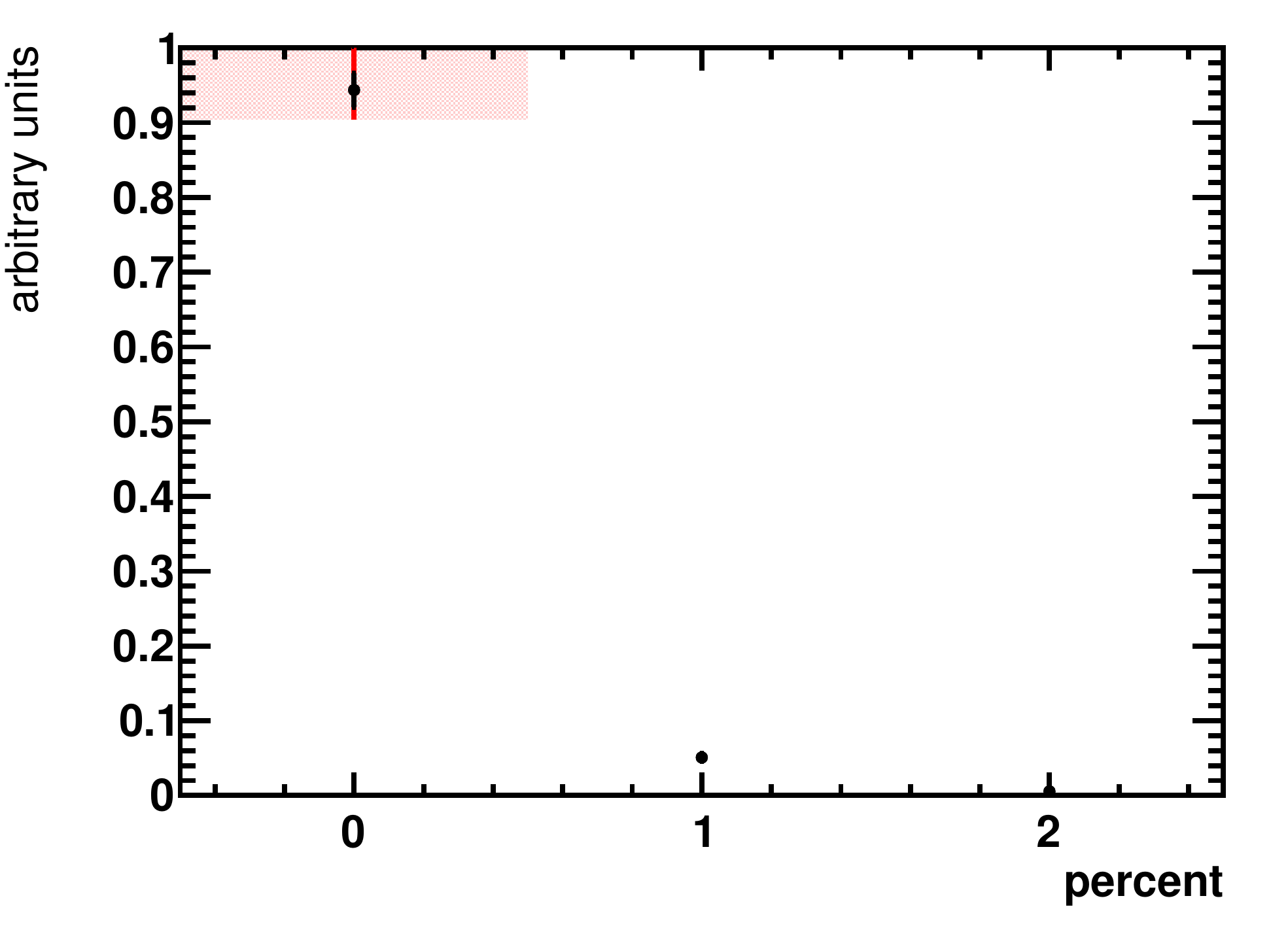}
\includegraphics[width=1.5in]{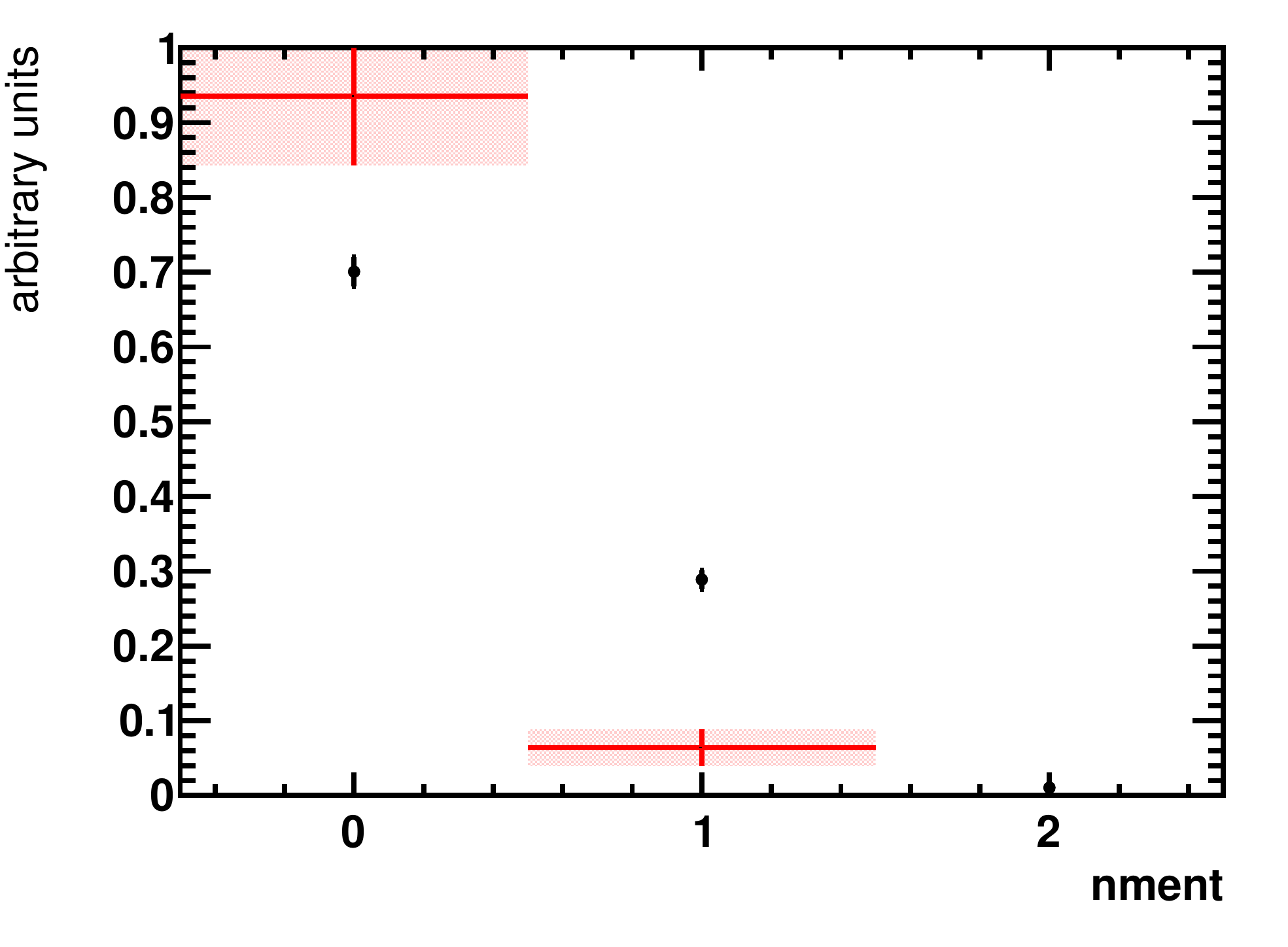}
\includegraphics[width=1.5in]{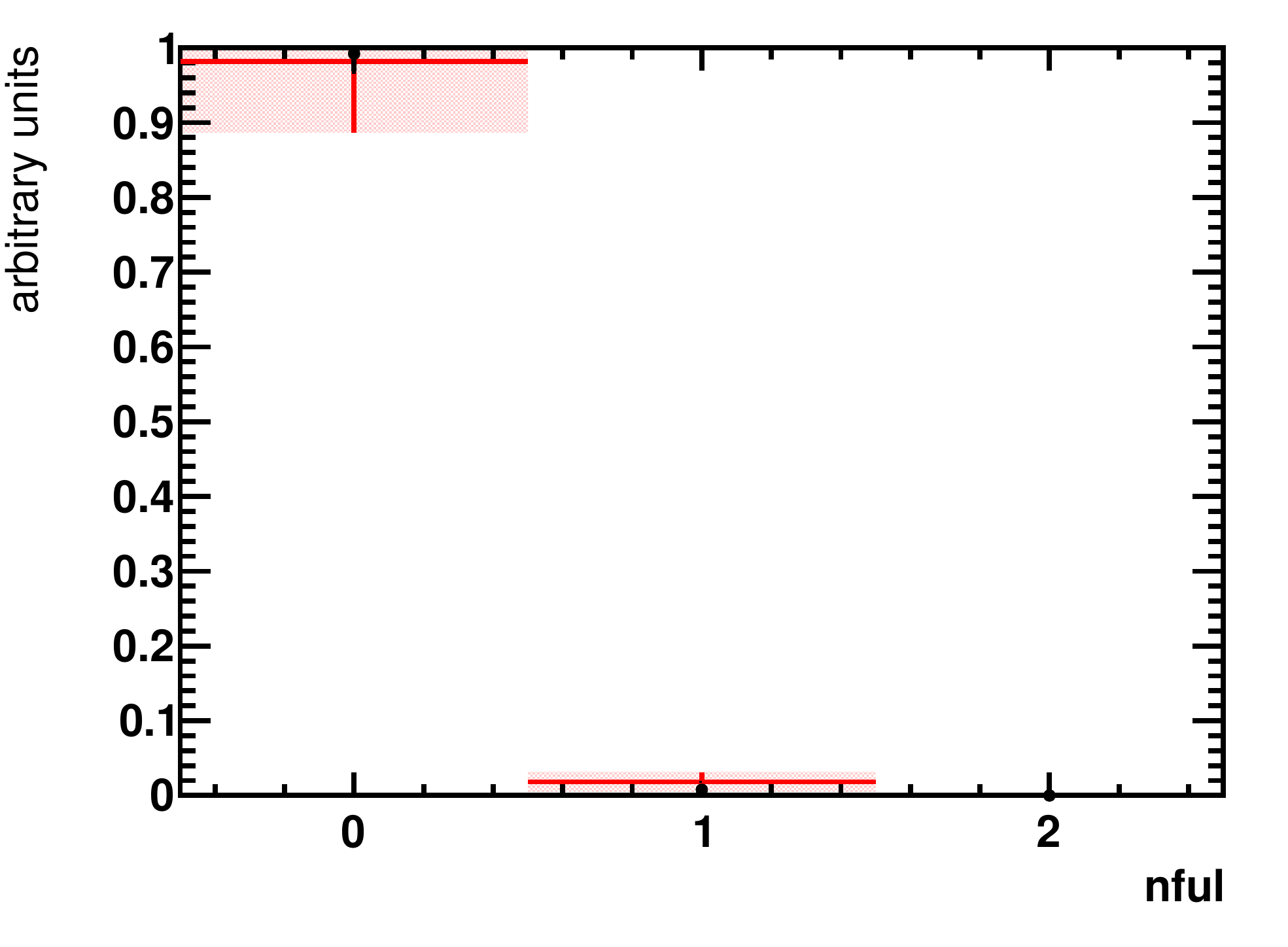}
\includegraphics[width=1.5in]{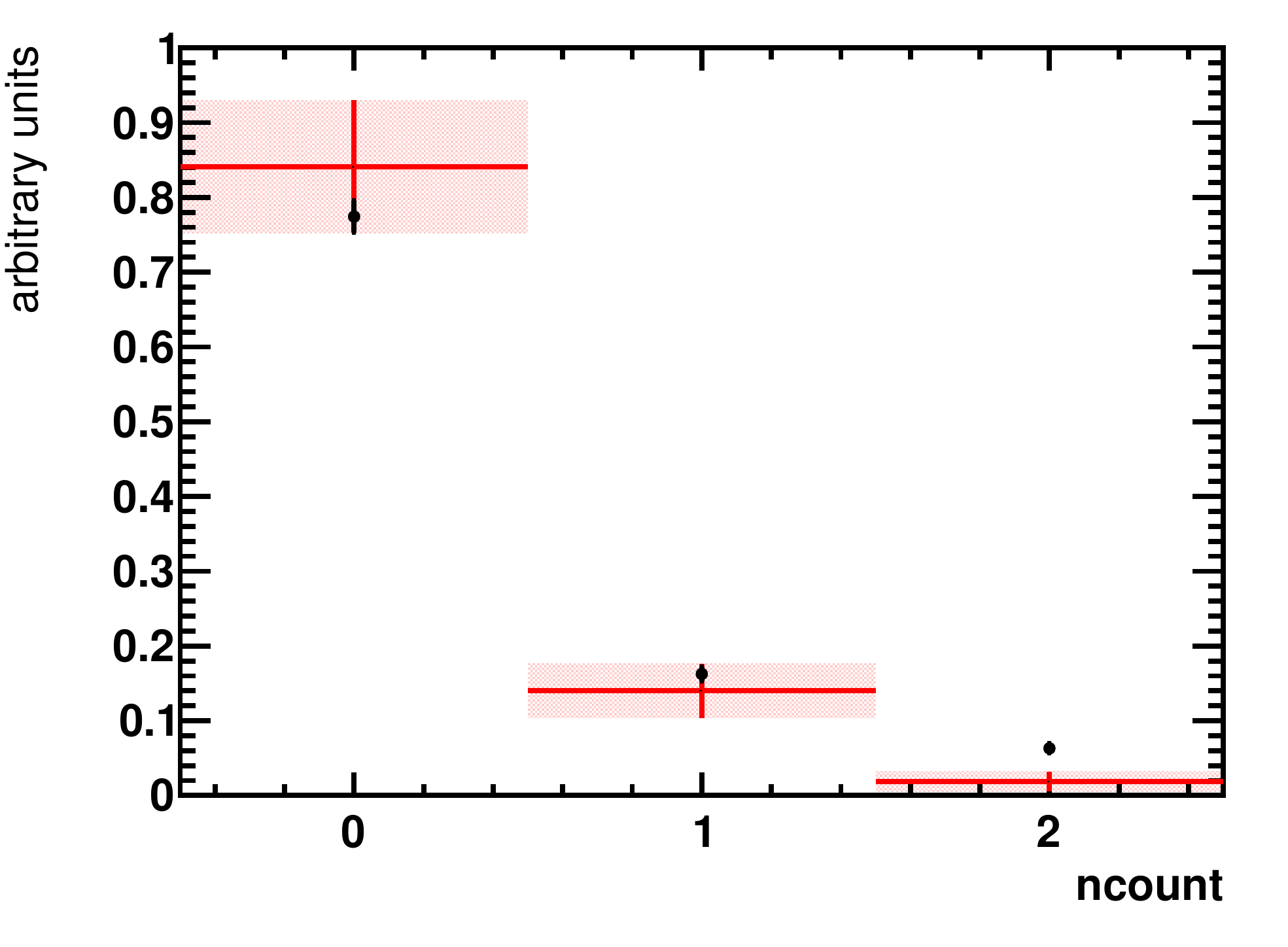}
\includegraphics[width=1.5in]{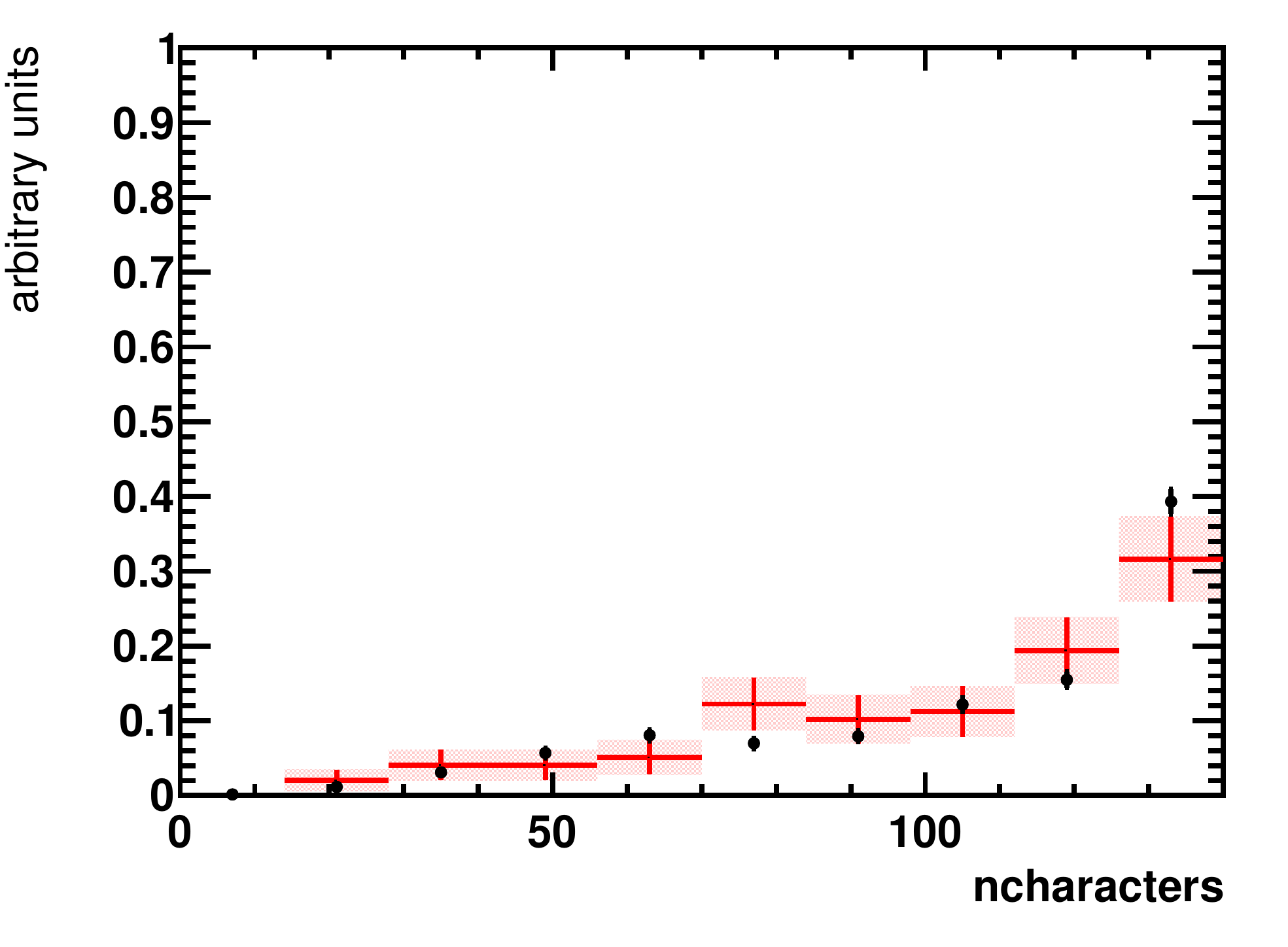}
\includegraphics[width=1.5in]{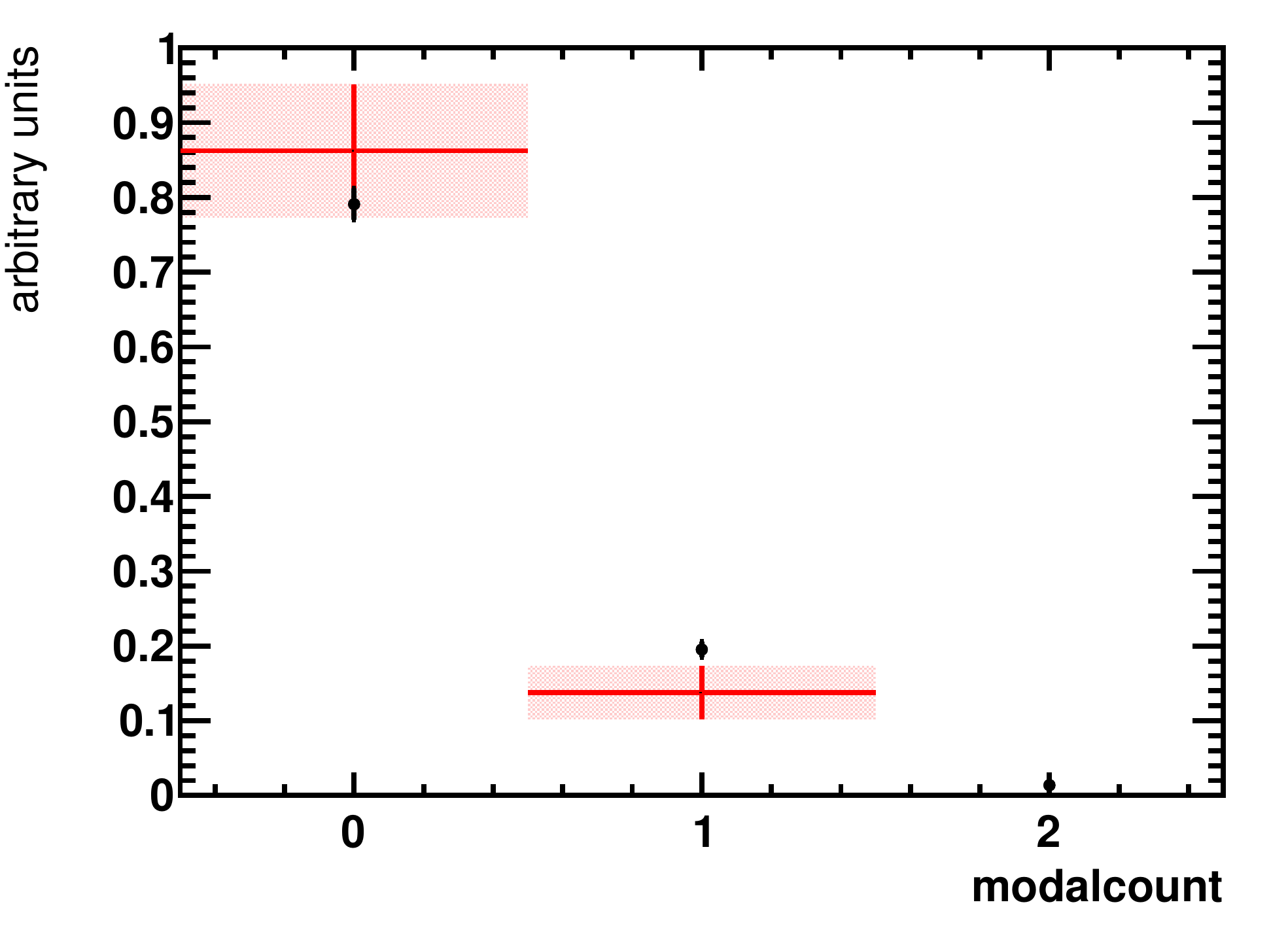}
\includegraphics[width=1.5in]{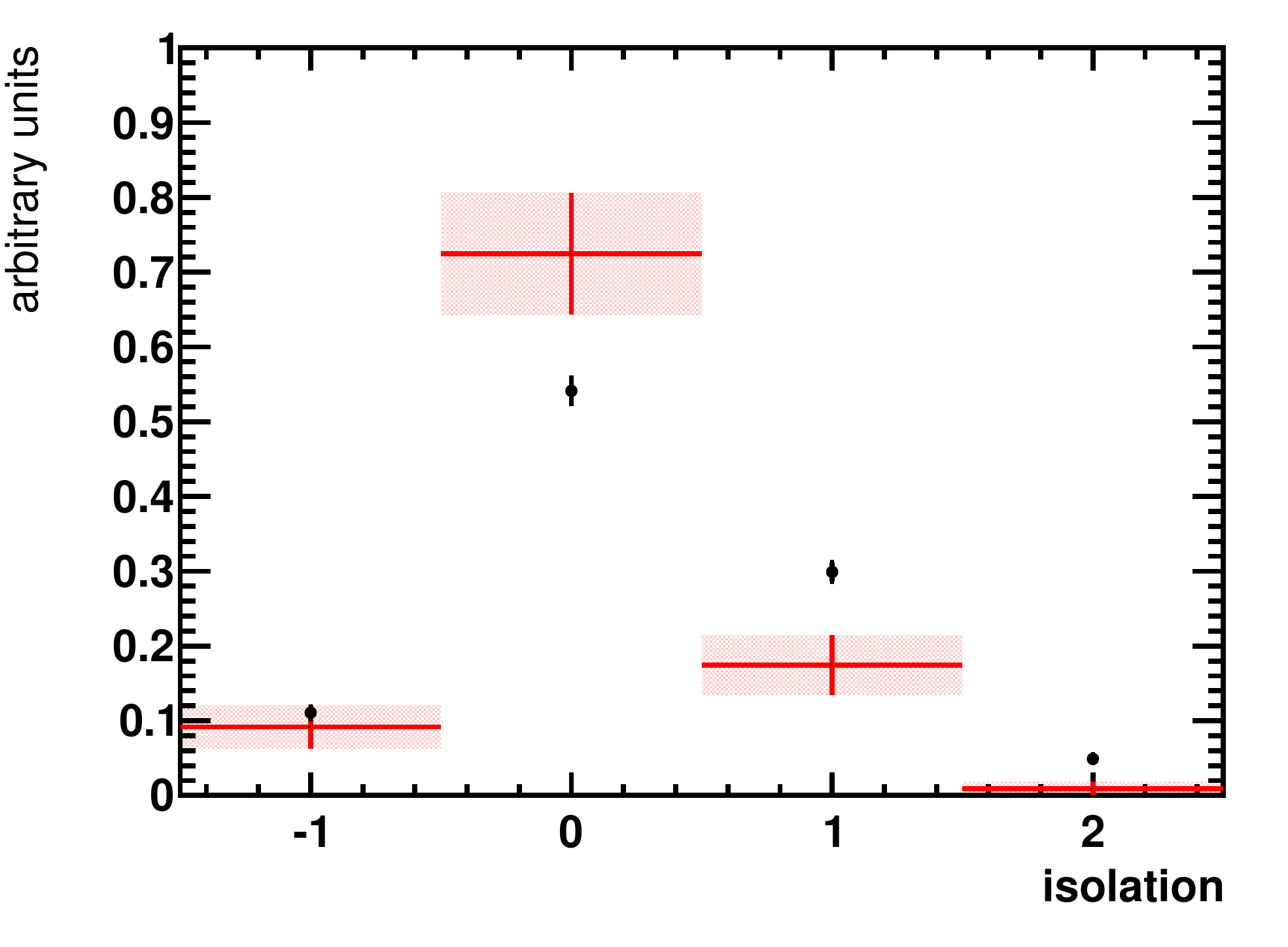}
\includegraphics[width=1.5in]{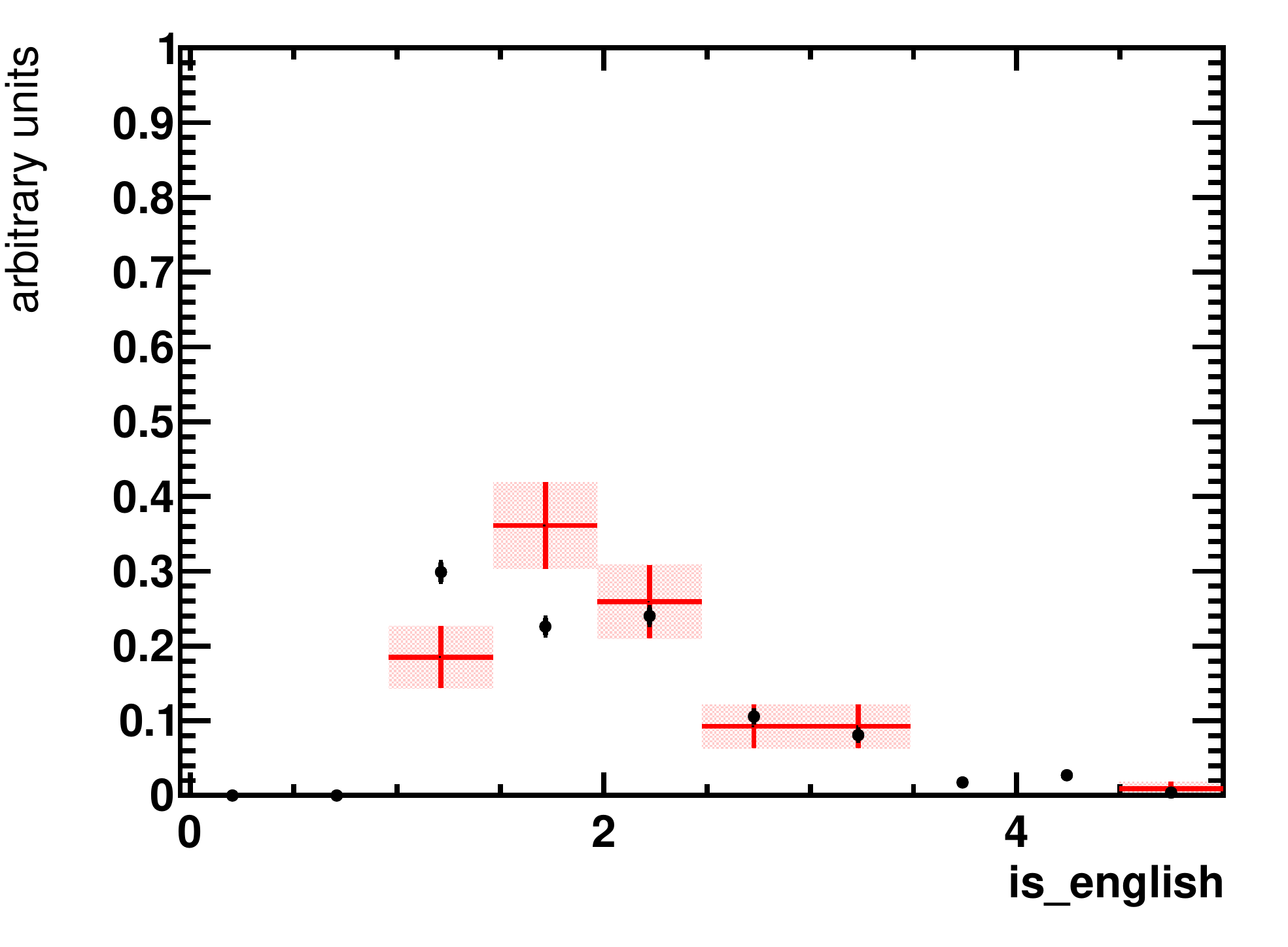}
\includegraphics[width=1.5in]{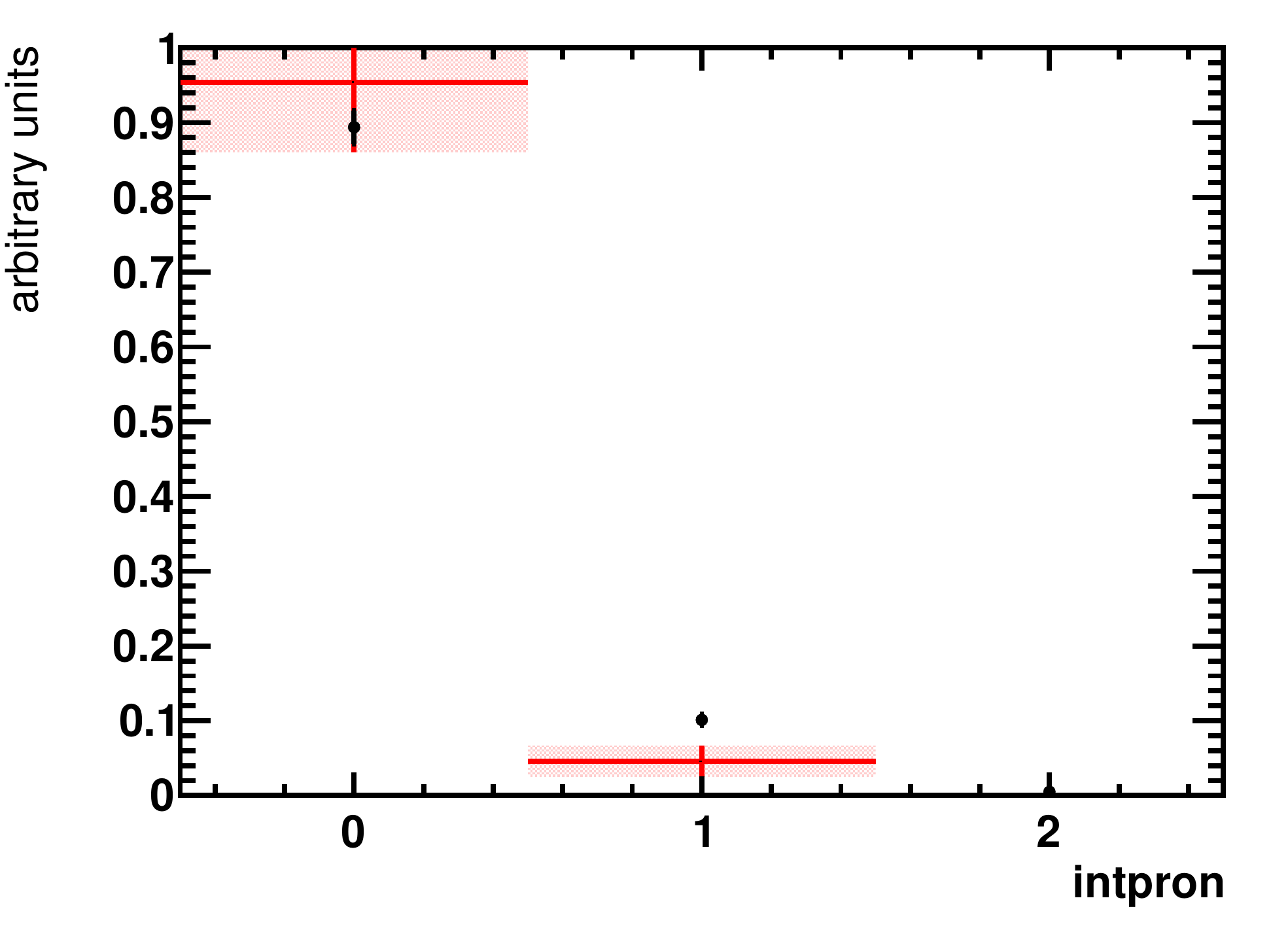}
\includegraphics[width=1.5in]{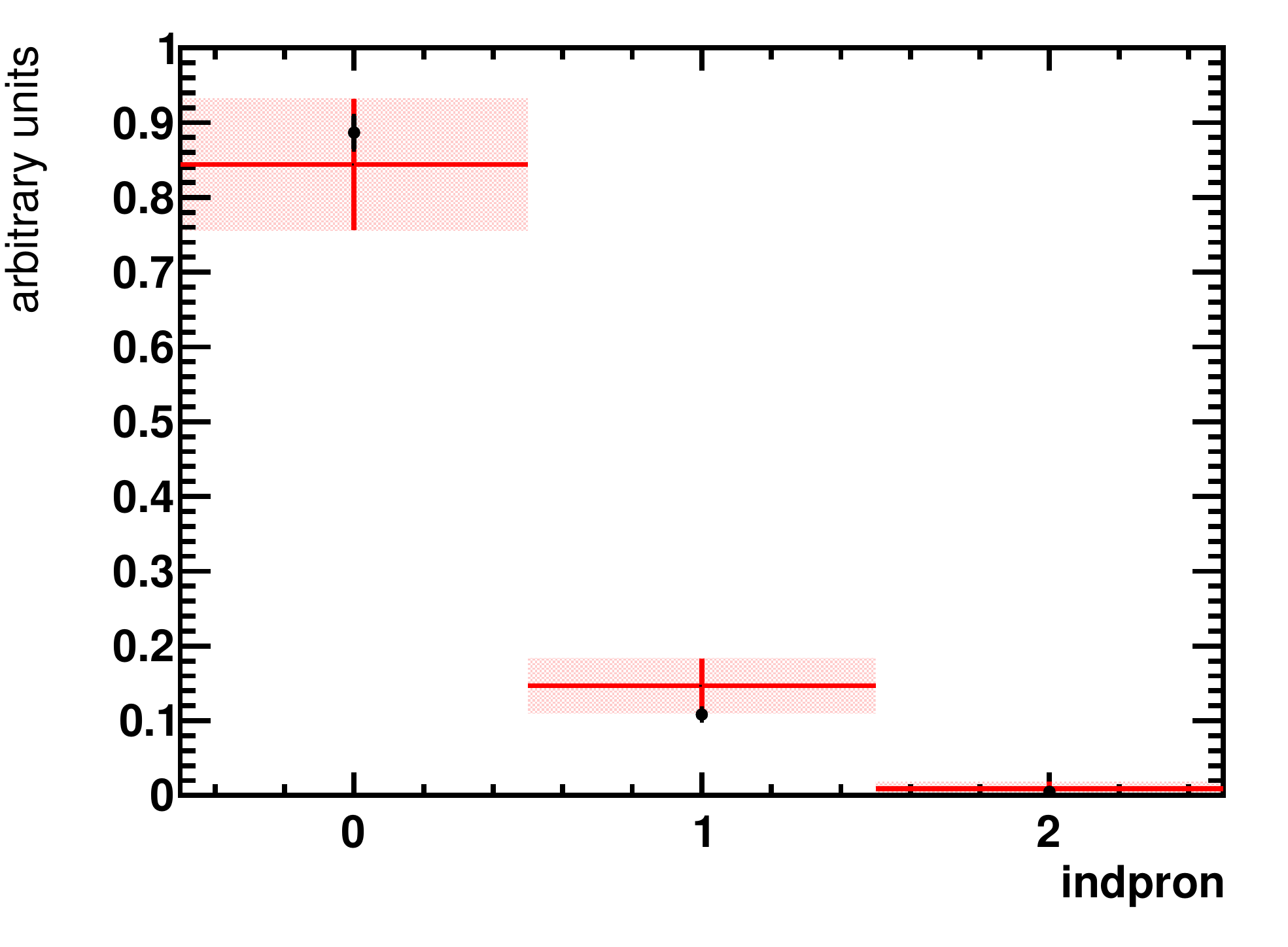}
\includegraphics[width=1.5in]{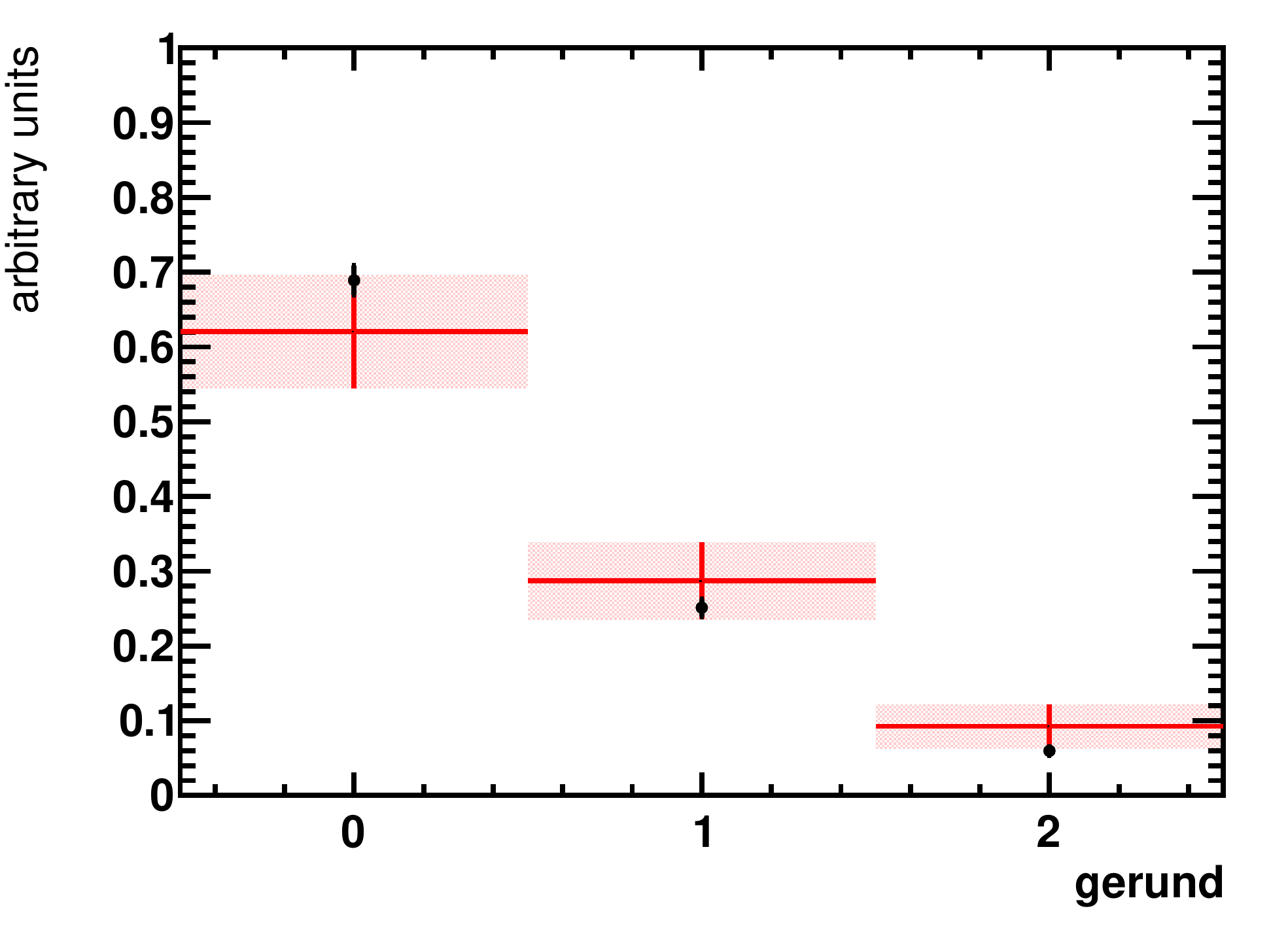}
\includegraphics[width=1.5in]{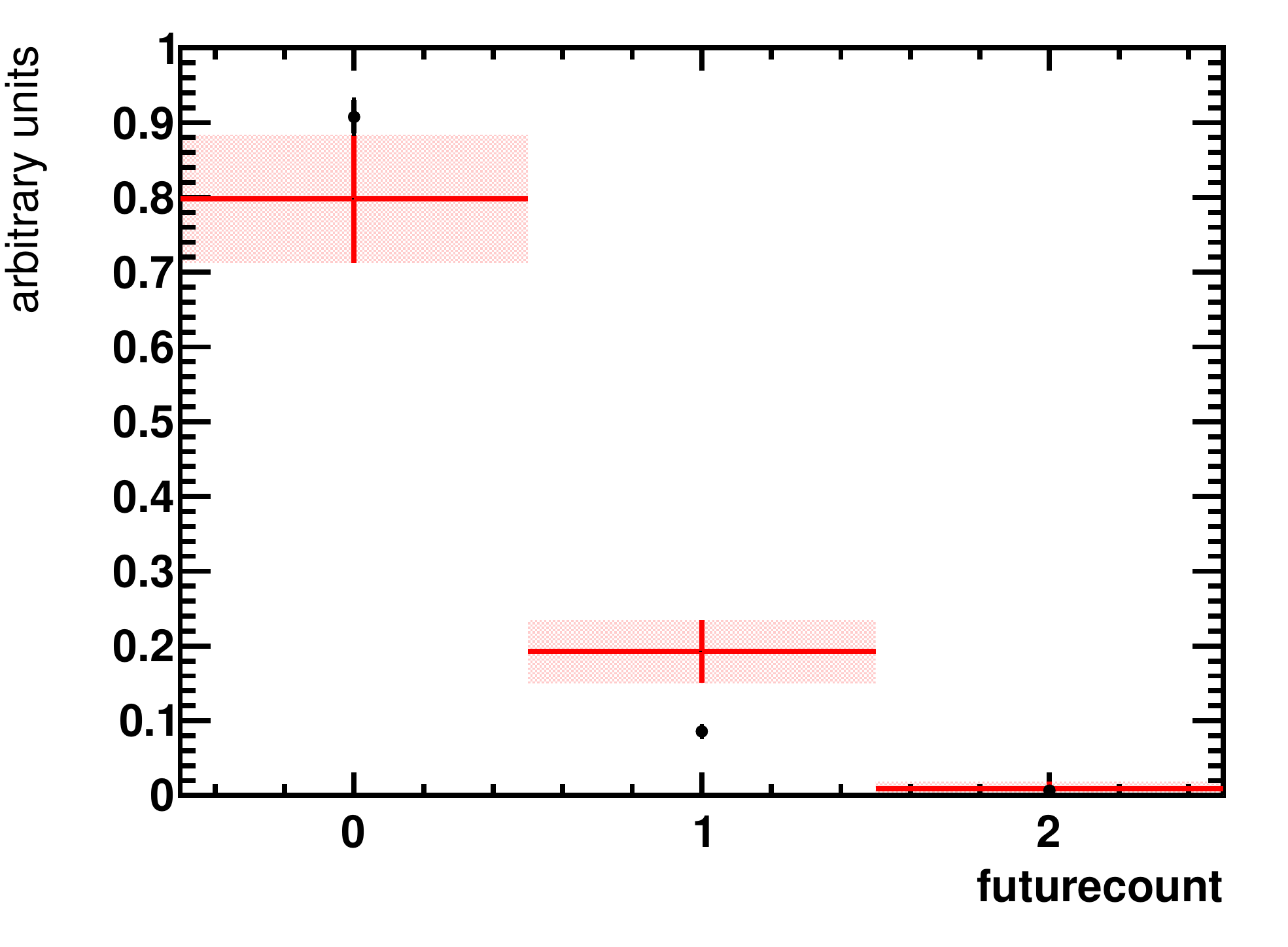}
\includegraphics[width=1.5in]{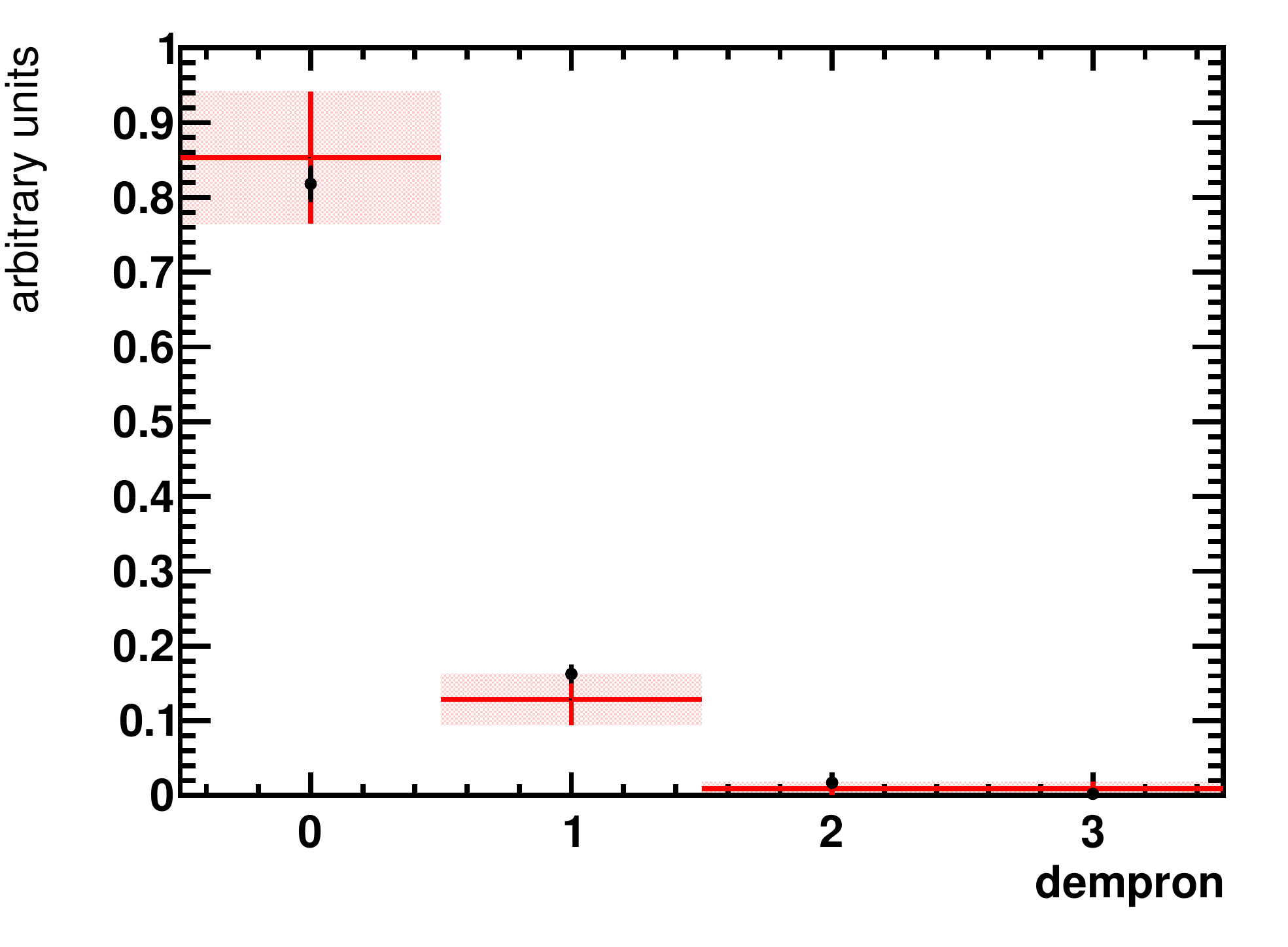}
\includegraphics[width=1.5in]{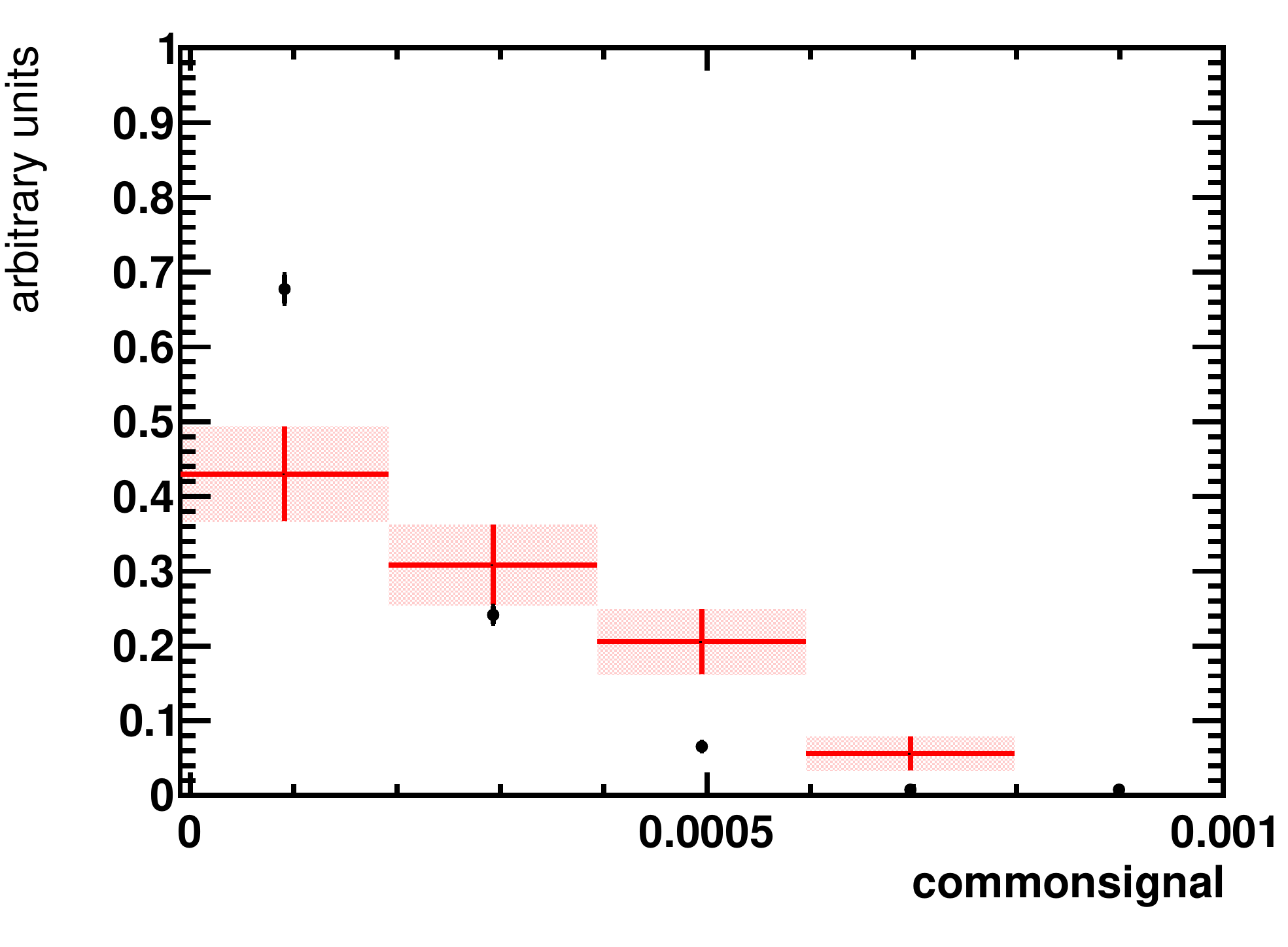}
\includegraphics[width=1.5in]{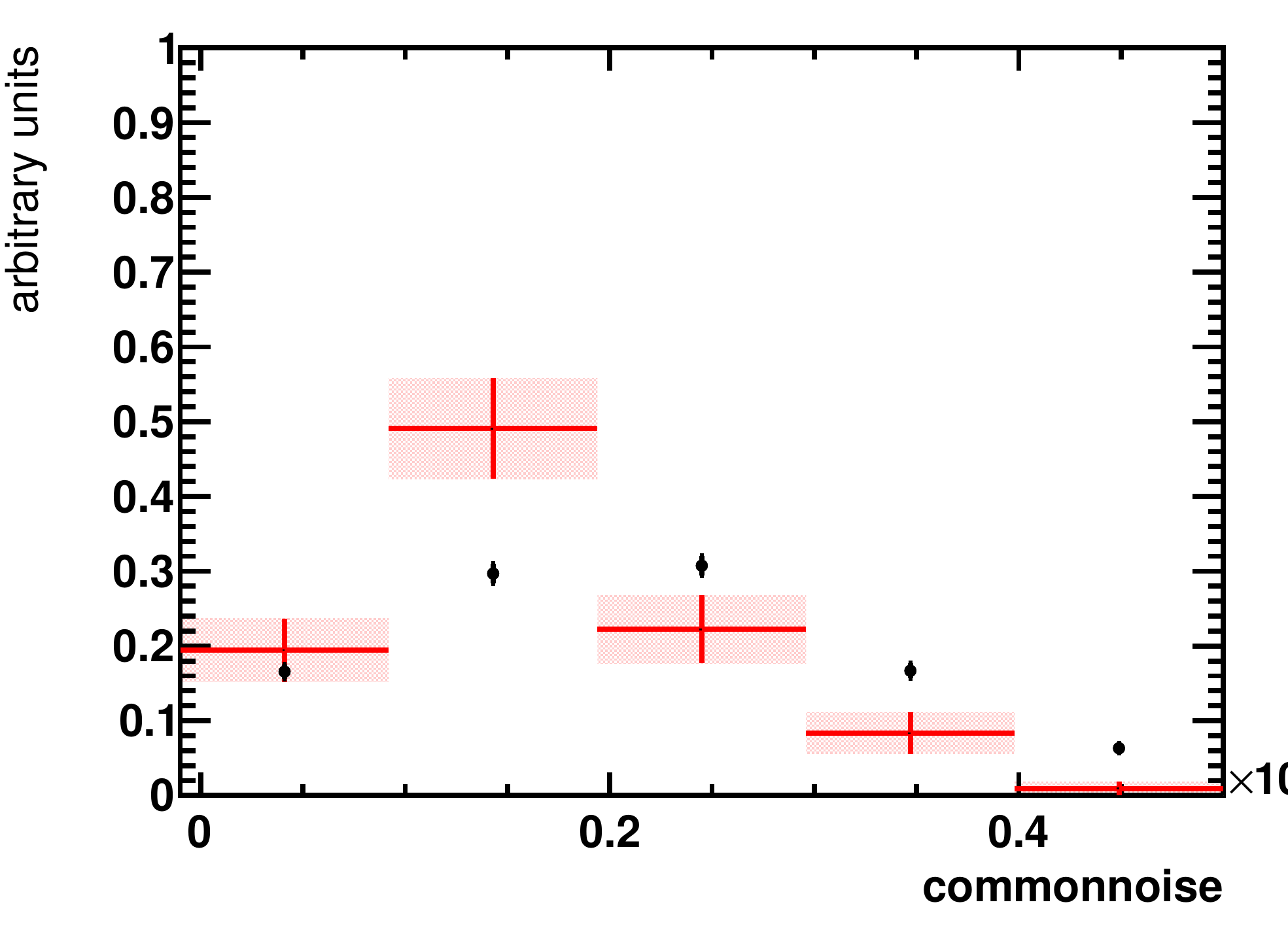}
\includegraphics[width=1.5in]{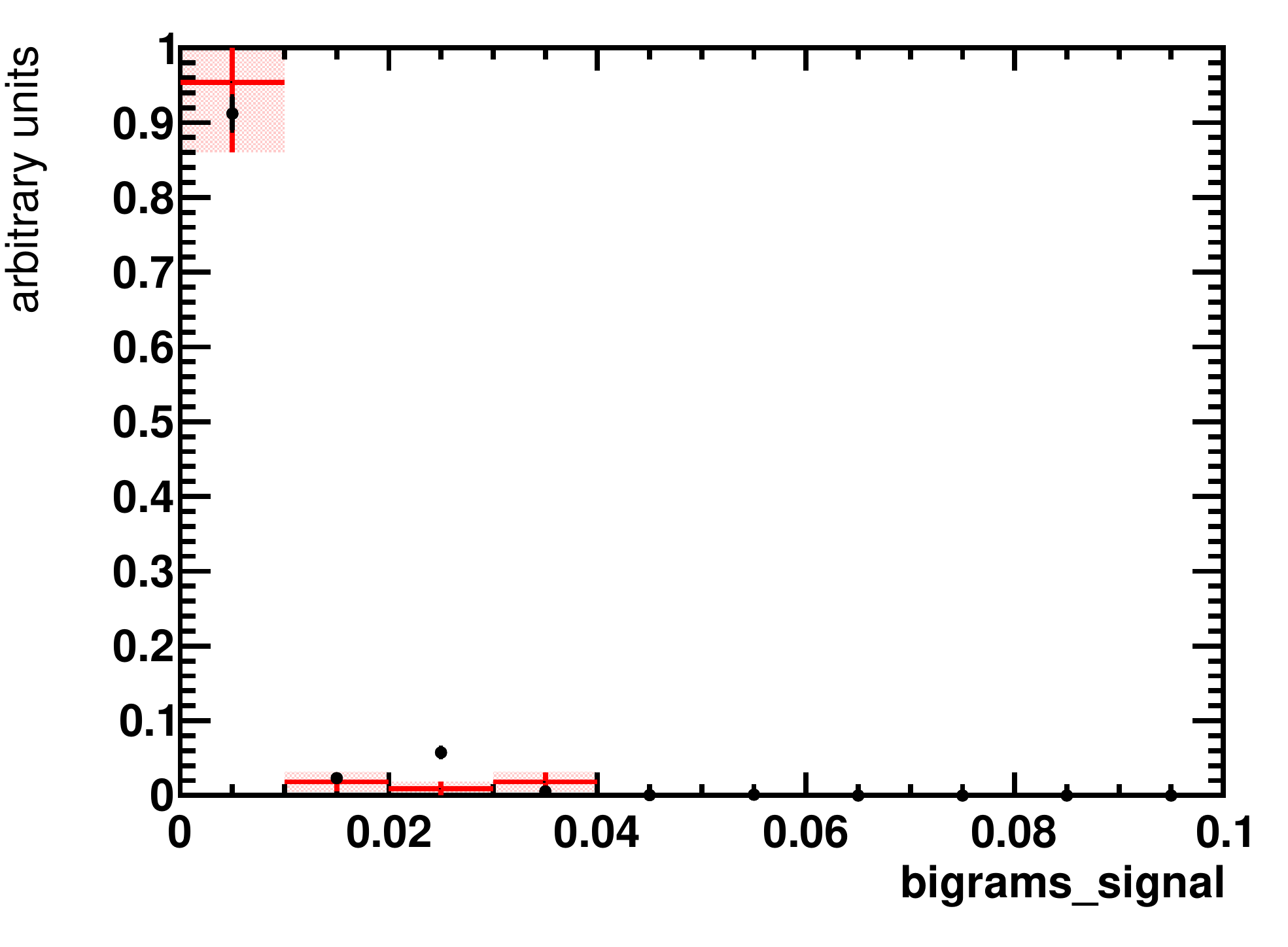}
\includegraphics[width=1.5in]{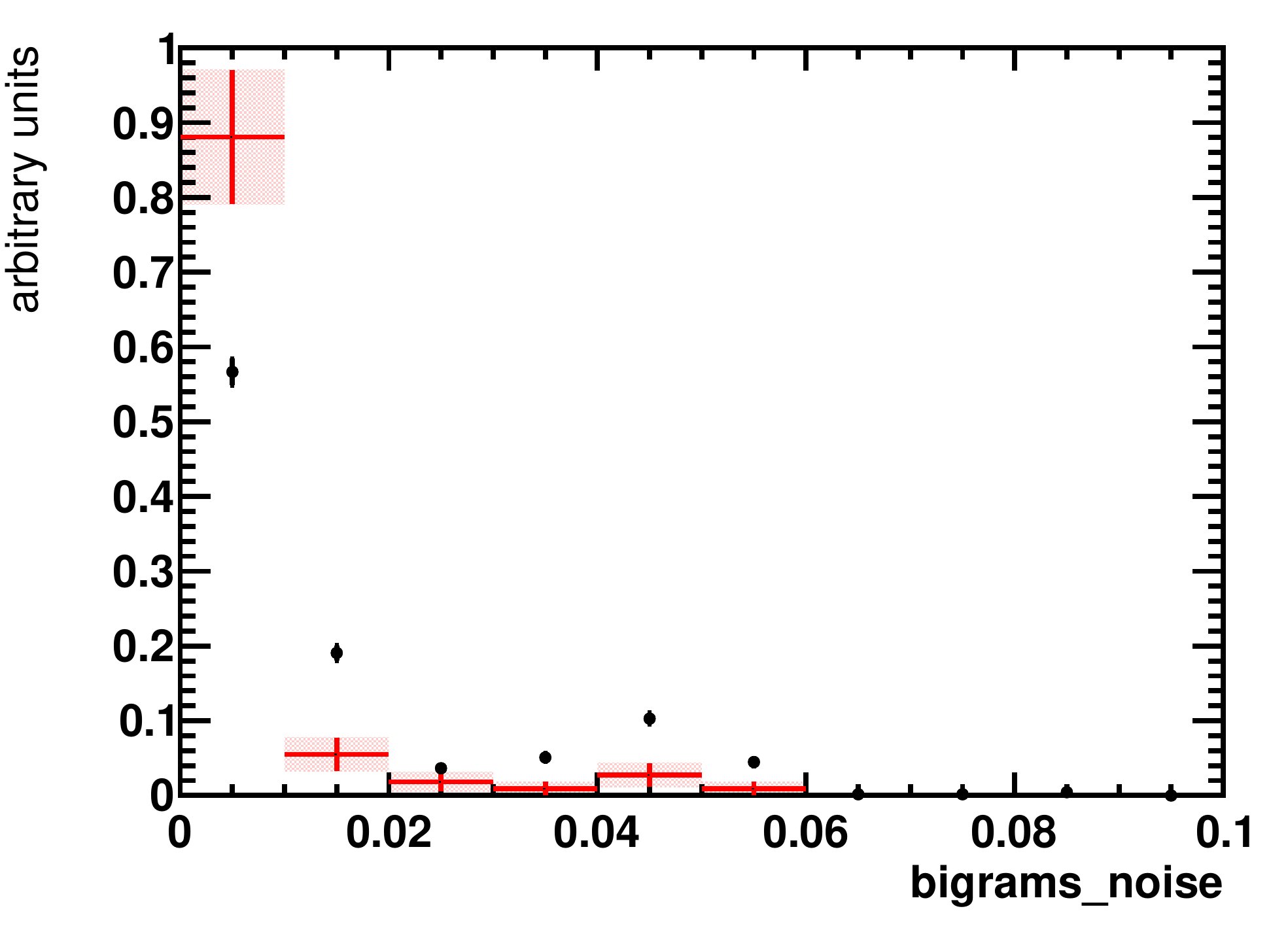}
\caption{First set of features for signal (red) and noise (black). The statistical uncertainties are represented on the vertical axis.} 
\label{hiv:variables:1}
\end{figure*}

\begin{table}\label{featuresimportance}
\centering
\caption{Features ranked by its separation power between signal and noise.}
\begin{tabular}{|l|l|l|}
\hline
 Rank & Variable       & Separation \\
\hline
\hline
    1 & personalcount  & 2.853$\times 10^{-01}$ \\
    2 & bigrams\_noise  & 2.689$\times 10^{-01}$ \\
    3 & posnoise       & 2.389$\times 10^{-01}$ \\
    4 & tagnoun        & 2.339$\times 10^{-01}$ \\
    5 & sis\_noise      & 2.182$\times 10^{-01}$ \\
    6 & ncharacters    & 1.971$\times 10^{-01}$ \\
    7 & commonnoise    & 1.727$\times 10^{-01}$ \\
    8 & tagconj        & 1.641$\times 10^{-01}$ \\
    9 & wordcount      & 1.549$\times 10^{-01}$ \\
   10 & sis\_signal     & 1.489$\times 10^{-01}$ \\
   11 & tagdeterm      & 1.442$\times 10^{-01}$ \\
   12 & tagadv         & 1.438$\times 10^{-01}$ \\
   13 & tagprep        & 1.379$\times 10^{-01}$ \\
   14 & commonsignal   & 1.350$\times 10^{-01}$ \\
   15 & tagadj         & 1.229$\times 10^{-01}$ \\
   16 & tagverb        & 1.138$\times 10^{-01}$ \\
   17 & tagto          & 1.044$\times 10^{-01}$ \\
   18 & is\_english     & 1.033$\times 10^{-01}$ \\
   19 & in\_notenglish  & 7.112$\times 10^{-02}$ \\
   20 & nment          & 6.890$\times 10^{-02}$ \\
   21 & secondpron     & 6.648$\times 10^{-02}$ \\
   22 & isolation      & 6.029$\times 10^{-02}$ \\
   23 & regularpast    & 4.609$\times 10^{-02}$ \\
   24 & bigrams\_signal & 4.309$\times 10^{-02}$ \\
   25 & modalcount     & 4.155$\times 10^{-02}$ \\
   26 & ncount         & 3.524$\times 10^{-02}$ \\
   27 & relatpron      & 3.280$\times 10^{-02}$ \\
   28 & thirdpron      & 3.081$\times 10^{-02}$ \\
   29 & gerund         & 2.707$\times 10^{-02}$ \\
   30 & percent        & 2.656$\times 10^{-02}$ \\
   31 & dempron        & 1.909$\times 10^{-02}$ \\
   32 & intpron        & 1.646$\times 10^{-02}$ \\
   33 & negative       & 1.283$\times 10^{-02}$ \\
   34 & pharmacy       & 1.093$\times 10^{-02}$ \\
   35 & indpron        & 6.787$\times 10^{-03}$ \\
   36 & nful           & 5.266$\times 10^{-03}$ \\
   37 & futurecount    & 5.259$\times 10^{-03}$ \\
\hline
\end{tabular}
\end{table}

\pagebreak

\subsection*{Foreign Language Removal}
We extracted features from tweets to be able to feed a machine learning algorithm, and separate noise from signal
more efficiently. Nevertheless, even if we had the best semantic features, the machinery would have difficulties in separating tweets that are not in English. In this section we propose a method to suppress almost all
of foreign tweets without losing much signal.

\begin{figure}
\centering
\caption{Distributions of is\_english for tweets annotated as not English (blue with black dots) and the rest of tweets (red).}
\includegraphics[width=3in]{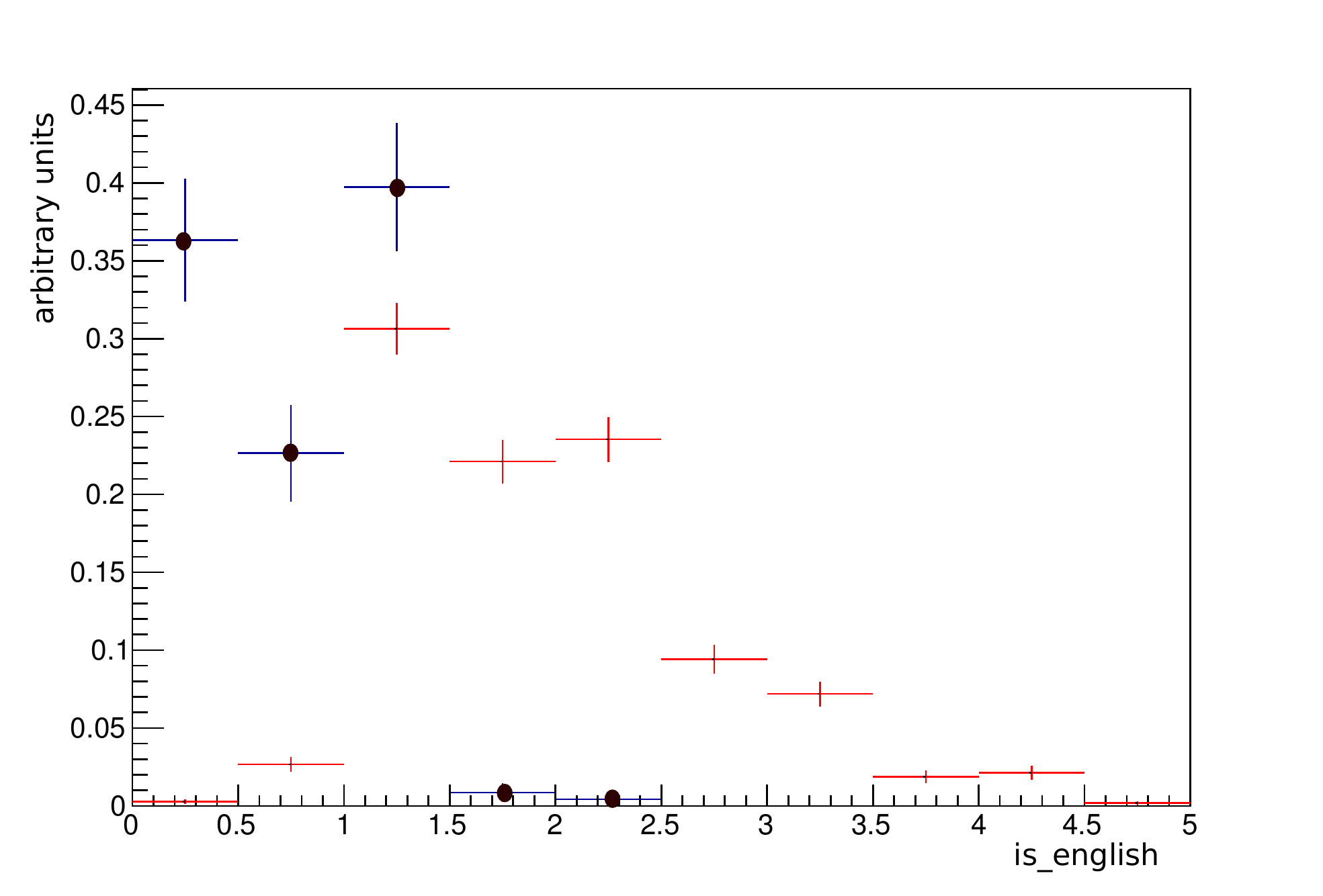}
\label{hiv:cuts:isenglish}
\end{figure}

\begin{figure}
\centering
\caption{Distributions of is\_english for signal (blue with black dots) and tweets annotated as English (red).}
\includegraphics[width=3in]{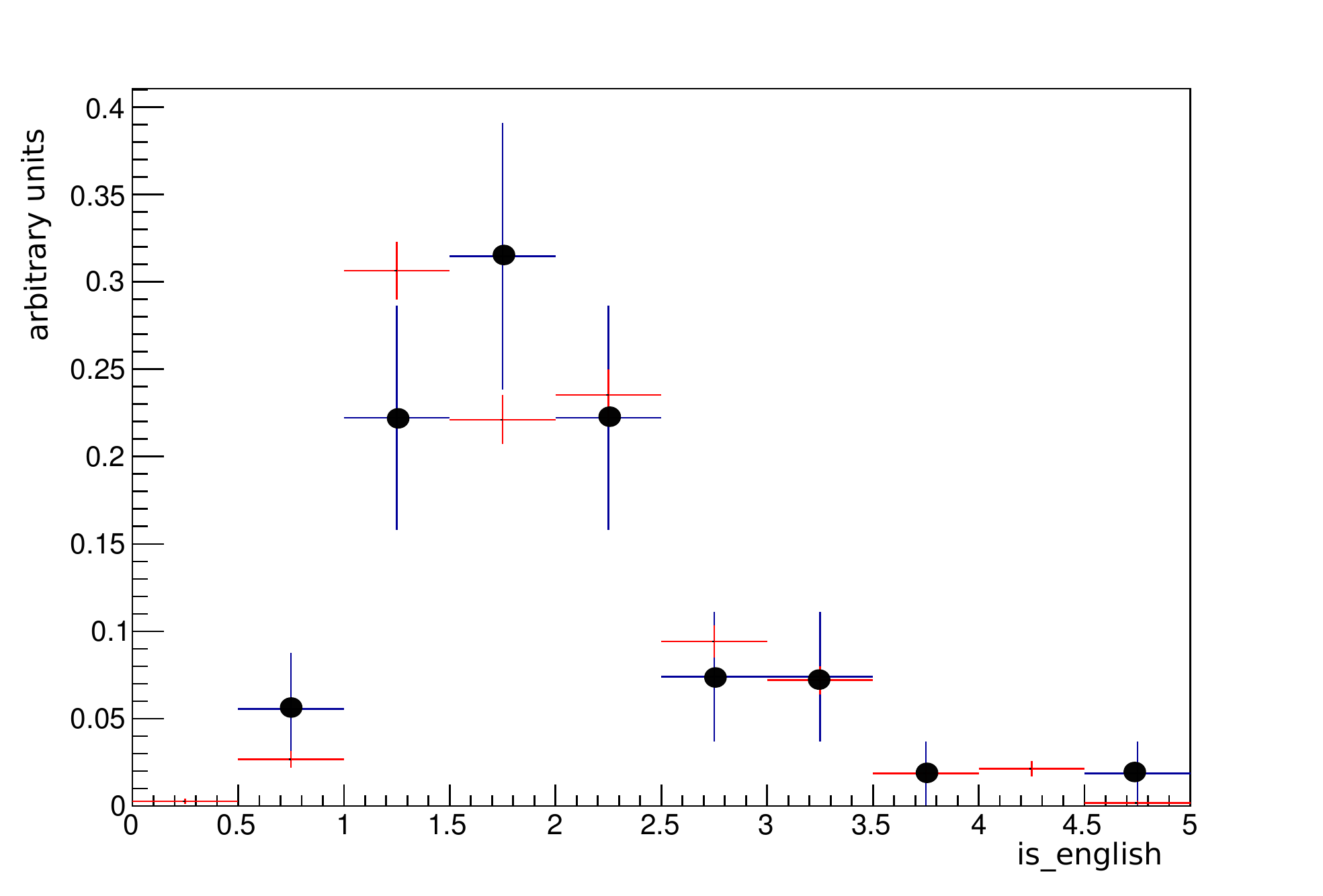}
\label{hiv:cuts:sigvseng}
\end{figure}

\vskip 2mm
We recall that we used tweets rated as not English as our control sample. Fig.~\ref{hiv:cuts:isenglish} shows the distributions of is\_english for tweets rated as not English and the rest. 60\% of not English tweets and 2.5\% of the rest remain at values of is\_english below 1. Therefore, we would reject 2.5\% of signal tweets and 60\% of not English tweets if we required is\_english$\geq$1.
The assumption that the distribution of\\is\_english for signal tweets and English tweets is the same, is validated by Fig.~\ref{hiv:cuts:sigvseng}.

\begin{figure}
\centering
\caption{Distributions of in\_notenglish for tweets annotated as not English (blue with black dots) and the rest of tweets (red).}
\includegraphics[width=3in]{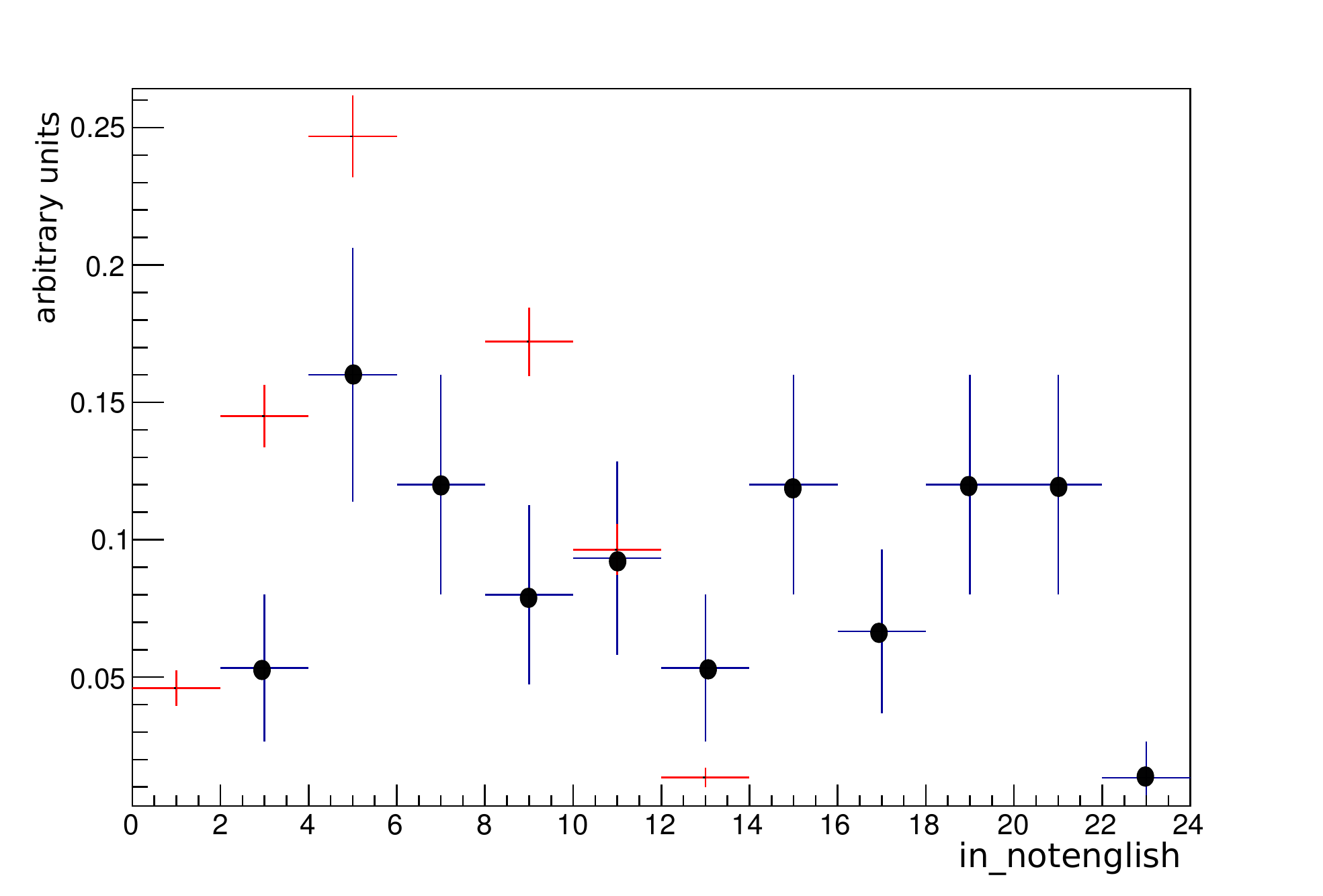}
\label{hiv:cuts:innotenglish}
\end{figure}

%\vskip 2mm
%Taking words from a non-english rated corpus, I define in\_notenglish as the number of words in the tweet that are also in the non-english list (Fig.~\ref{hiv:cuts:innotenglish}).
Fig.~\ref{hiv:cuts:innotenglish} shows the distributions of in\_notenglish for tweets annotated as not English and the rest of annotated tweets.

\begin{figure}
\centering
\caption{$\frac{wordcount}{in\_notenglish}$ for tweets annotated as not English (blue with black dots) and the rest of annotated tweets (red).}
\includegraphics[width=3in]{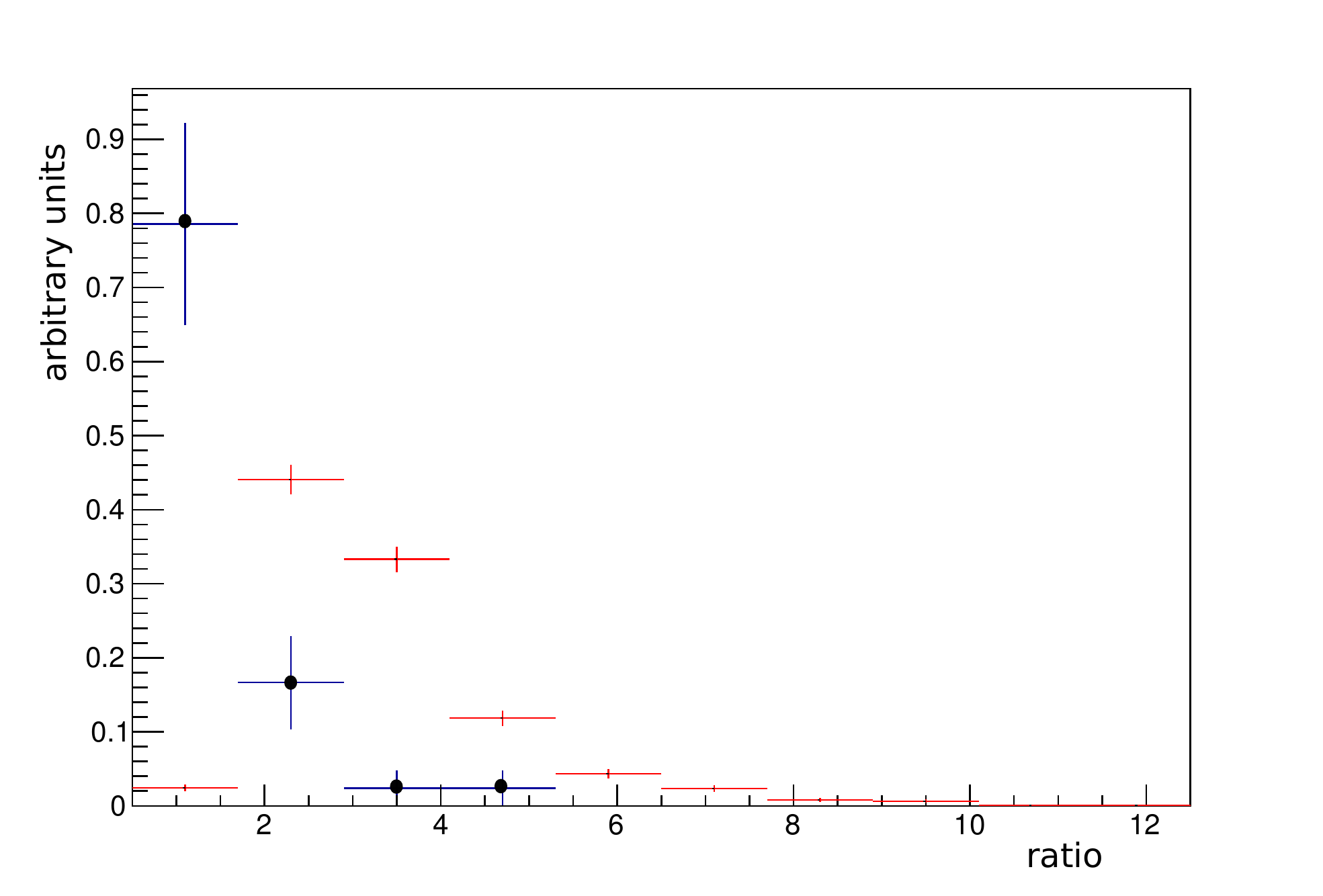}
\label{hiv:cuts:ratio}
\end{figure}

Almost all values of in\_notenglish are equal to wordcount for tweets annotated as not English. Fig.~\ref{hiv:cuts:ratio} shows the ratio of wordcount by in\_notenglish for tweets annotated as not English and the rest of tweets. 

In conclusion, Tab.~6 summarizes the yields before and after the requirements: is\_english$\ge$1, ncharacters$<$150,\\in\_notenglish$<$14, $\frac{wordcount}{in\_notenglish}>1$. These requirements lead to an estimated 6$\%$ signal lose, while removing 20$\%$ of all noise and 94$\%$ of not English tweets. 

\begin{table}\label{hiv:cuts:table}
\caption{Yields before and after requirements.}
\centering
\begin{tabular}{|l|l|l|}
\hline
Type & Yield before & Yield after \\
\hline
\hline
%Signal & 54 & 50\\
%Noise (all) & 1380 & 1107 \\
%Noise (Not English) & 234 & 13\\
%\hline
%\hline
%Signal & 40 & 38\\
%Noise (all) & 1337 & 1072 \\
%Noise (Not English) & 238 & 16\\
Signal & 94 & 88 \\
Noise (all) & 2717 & 2179 \\
Noise (not English) & 472 & 29 \\
\hline
\end{tabular}
\end{table}

\pagebreak

\subsection*{Machine learning classifier}
The goal of the data curation described through this appendix is to reduce the original sample of tweets to a sample containing only signal. To define training samples for signal, noise and not English, we performed a crowdsourcing request of 2,000 tweets. These tweets were rated by two Amazon Mechanical Turk workers. We believed that another request of about 5,000 $\pm$ 1,000 tweets to be annotated would not compromise our budget for future studies. Therefore, our goal was indeed to reduce the datasample to the quoted 5,000 $\pm$ 1,000 tweets. This implied a reduction of the noise of the order of 80$\%$. Hereafter we detail the steps that we took to reach such reduction.

\begin{figure}
\centering
\caption{Signal efficiency versus (Background) noise rejection for four classifiers: Boosted Decision Trees with AdaBoost (BDT), Support Vector Machines (SVM), Boosted Decision Trees with Bagging (BDTB), and Artificial Neural Networks (MLP). The classifiers are defined using all features shown in Tab.~5.}
\includegraphics[width=3in]{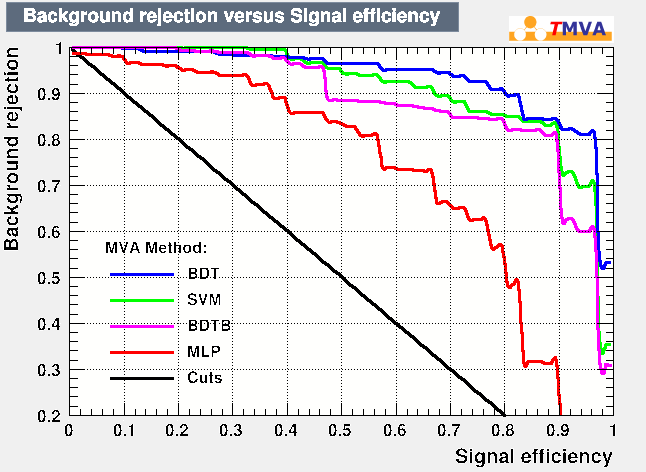}
\label{hiv:ml:rocall38}
\end{figure}

\vskip 2mm
We used TMVA~\cite{tmva} library to compute a machine learning classifier using all features described above. We split our samples according to Tab.~2, and computed the signal efficiency versus noise rejection for four types of classifiers: Boosted Decision Trees with AdaBoost (BDT), Support Vector Machines (SVM), Boosted Decision Trees with Bagging (BDTG), and Artificial Neural Networks (MLP). The results are shown in Fig.~\ref{hiv:ml:rocall38}. This figure is a ROC rotated anti-clockwise ninety degrees. 

\begin{figure}
\centering
\caption{Signal efficiency versus (Background) noise rejection for four classifiers: Boosted Decision Trees with AdaBoost (BDT), Support Vector Machines (SVM), Boosted Decision Trees with Bagging (BDTB), and Artificial Neural Networks (MLP). The classifiers are defined using personalcount, bigrams\_noise, tagnoun, sis\_noise, ncharacters, commonnoise, commonsignal, is\_english, sis\_signal.}
\includegraphics[width=3in]{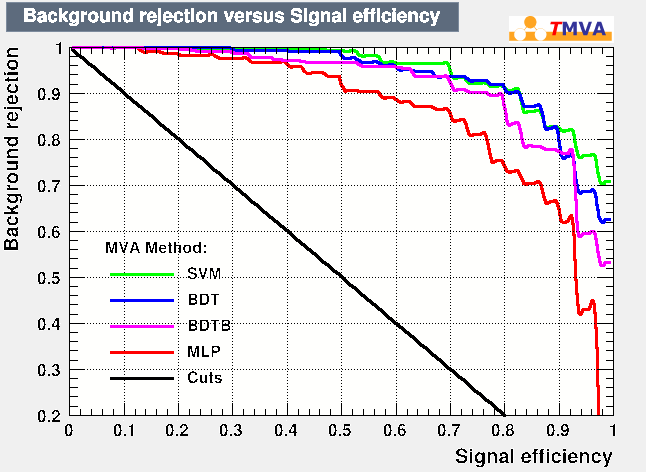}
\label{rocallfinal}
\end{figure}

\vskip 2mm
Tab.~5 indicates that the features at the top of the ranking have quantifiably larger separation power than those at the bottom. We tested, step by step, whether removing less performant features would have an impact on the overall performance of the classifier. At the end of this procedure that consisted in removing variables, a set of nine variables was chosen. This decision was taken given the inherent simplicity of the classifier defined with fewer variables and the similar separation power between signal and noise of the mentioned classifier defined with less variables. Fig.~\ref{rocallfinal} shows the ROC obtained for the aforementioned machine learning algorithms, defined with the variables: personalcount, bigrams\_noise, tagnoun, sis\_noise, ncharacters, commonnoise, commonsignal, is\_english, sis\_signal.

\vskip 2mm
In Fig.~\ref{rocallfinal}, the SVM classifier shows the best performances in our region of interest. Moreover, this classifier shows less dependency on the number of variables than BDT. We chose SVM trained with the aforementioned nine variables to define our classifier. Then, we compared the performances of our classifier with two different testing samples and verified a good compatibility (Fig.~\ref{comparison}).

\vskip 2mm
The output of our classifier is a real number between 0 and 1. Higher values indicate higher probability of being signal.
In order to estimate the threshold to apply to our sample, we used the annotated signal tweets and computed the 90\% signal efficiency threshold to be 0.45. Therefore, we parsed our entire sample of tweets through our classifier and kept only those tweets with classifier outputs larger than 0.45. As a result of this filtering, our remaining sample was reduced to 5443 tweets that we sent for crowdsourcing rating. Finally, after crowdsourcing, our pure sample of signal contained 1642 annotated tweets.

\end{appendix}

\end{document}